\documentclass{aa}    
    
\usepackage{graphicx}
\usepackage{txfonts}
\usepackage{natbib} 
\usepackage{subfigure}
\usepackage{color}
\usepackage{subeqnarray}
\usepackage{natbib} 
\usepackage{version}
\usepackage{rotating}

\begin{document}

\def\be{\begin{equation}}
\def\ee{\end{equation}}
\def\bea{\begin{eqnarray}}
\def\eea{\end{eqnarray}}
\def\ttb{\textbf}

\def\Sun{\odot}
\def\um{$\mu$m}
\def\cl{$C_\ell$}

\def\be{\begin{equation}}
\def\ee{\end{equation}}
\def\bea{\begin{eqnarray}}
\def\eea{\end{eqnarray}}
\def\ttb{\textbf}

\title{Modeling the evolution of infrared galaxies~: clustering of galaxies in the Cosmic Infrared Background}
\author{Aur\'elie P\'enin \inst{1,2}
\and Olivier Dor\'e\inst{3,4}
\and Guilaine Lagache\inst{1,2}
\and Matthieu B\'ethermin\inst{1,2}}
\institute{Univ. Paris Sud, Institut d'Astrophysique Spatiale, UMR8617, F-91405 Orsay
\and CNRS F-91405 Orsay 
\and Jet Propulsion Laboratory, California Institute of Technology, Pasadena, CA 91109
\and California Institute of Technology, Pasadena, CA 91125}


\abstract{Star-forming galaxies are a highly biased tracer of the underlying dark matter density field. Their clustering can be studied through the cosmic infrared background anisotropies. These anisotropies have been measured from 100 \um~to 2 mm in the last few years.}
{In this paper, we present a fully parametric model allowing a joint analysis of these recent observations.}
{In order to develop a coherent model at various wavelengths, we rely on two building blocks. The first one is a parametric model that describes the redshift evolution of the luminosity function of star-forming galaxies. It was introduced in B\'ethermin et al. (2011) and compares favorably to measured differential number counts and luminosity functions. The second one is a halo model based description of the clustering of galaxies. Starting from a fiducial model, we investigate parameter  degeneracies using a Fisher analysis. We then discuss how halo of different mass and redshift, how LIRGs and ULIRGs, contribute to the CIB  angular power spectra.}
{From the Fisher analysis, we conclude that we cannot constrain the parameters of the model of evolution of galaxies using clustering data only. The use of combined data of \cl, counts and luminosity functions improves slightly the constraints but does not remove any degeneracies. On the contrary, the measurement of the anisotropies allows us to set interesting constraints on the halo model parameters, even if some strong degeneracies remain. Using our fiducial model, we establish that the 1-halo and 2-halo terms are not sensitive to the same mass regime. We also illustrate how the 1-halo term can be misinterpreted with the Poisson noise term.}
{We present a new model of the clustering of infrared galaxies. Our framework allows a coherent and joint analysis of various probes of infrared galaxies~: number counts, luminosity functions and clustering measurements. However such a model has a few limitations, as the parameters of the halo occupation suffer from strong degeneracies.}

\titlerunning{A parametric model of clustering}
\maketitle
%---------------------------------------------
\section{Introduction}
%---------------------------------------------

Infrared star-forming galaxies are mainly characterized by a very high star formation rate, tens or even hundreds times larger than that of the Milky Way, $\sim$10-100 $M_\odot$/year \citep{1998ApJ...498..541K}. The large number of young stars are embedded in dust that absorbs UV starlight and reemits it in the infrared (IR), from 5 $\mu$m to 1 mm. As a result, infrared star-forming galaxies emit most of their energy ($\sim$95\%) in the IR. In the far IR ($\lambda>200 \mu$m) and sub-millimeter, observations are limited by confusion, small spatial scales are lost because of the poor angular resolution of the instruments. Thus we observe the Cosmic Infrared Background (CIB) \citep{1996A&A...308L...5P,1998ApJ...508..123F} which is the contribution of infrared star-forming galaxies integrated over the age of the Universe, and its anisotropies. In the near and mid IR a large fraction of the CIB is resolved into sources whereas in the far IR only a few percents is. These fractions can be improved thanks to the use of statistical methods. For instance, at long wavelength, \citet{2010AA...518L..21O} directly resolved 15\%, 10\% and 6\% of the CIB at 250, 350 and 500 $\mu$m in Herschel/SPIRE data whereas \citet{2004ApJS..154...70P} resolved 70\% of the CIB at 24~\um. \citet{2010AA...518L..30B} resolved 45\% and 52\% of the CIB at 100 and 160~$\mu$m respectively by integrating number counts in Herschel/PACS data. Stacking 24~$\mu$m sources they increased these fractions to 50\% and 75\% respectively. As a result, sources detected at 24~$\mu$m are the main sources of the CIB around its peak at $\sim$200 \um. The CIB is dominated by objects that get more and more massive as the redshift increases from luminous IR galaxies (LIRGs) with $10^{11}L_{\odot}<L_{IR}<10^{12}L_{\odot}$ at $0.8<z<1.2$ with intermediate mass, to extreme LIRGs with $10^{12}L_{\odot}<L_{IR}<10^{14}L_{\odot}$ that dominate at $z>2$ and with masses $> 10^{11}M_{\odot}$ \citep{2006astro.ph..4236C}.\\
\citet{2007MNRAS.375.1121M} and \citet{2006ApJ...641L..17F} derived the two-point correlation function of Ultra LIRGs at $z\simeq1.6-2.7$ and $1.5<z<3$ respectively. They brought to light the very strong clustering of infrared star-forming galaxies and their embedment in very massive halos of $\simeq10^{13}M_\Sun$. \citet{2010AA...518L..22C} computed the angular correlation function with Herschel/SPIRE data. They found that 250~$\mu$m sources are in DM halos with masses around $10^{12}M_\Sun$ that lie at $z\sim2.1$ whereas bright 500~$\mu$m sources are in more massive halos $\simeq10^{13}M_\Sun$ at $z\sim2.6$. More recently, \citet{2011MNRAS.tmp.1055M} derived the 3D correlation function of infrared sources using Herschel/PACS data up to a redshift of 3. They obtain that their galaxies lie in haloes with $>10^{12.4}M_\Sun$, value that is in agreement with previous studies. However the two-point correlation function is not easily computed using IR data because of confusion. As said before, confusion can be circumvented through the use of statistical methods. Indeed, clustering can be measured in the correlated CIB anisotropies (CIBA). It has first been detected as an excess of signal at intermediate scales by \citet{2007ApJ...665L..89L} and \citet{2007A&A...474..731G} at 160~$\mu$m in the Spitzer Multi-band Imaging Photometer (MIPS) data. These measurements have been followed by the detection in the Balloon-borne Large Aperture Sub-millimeter Telescope (BLAST) data at 250, 350 and 500~$\mu$m  \citep{2009ApJ...707.1766V} and by that of the South Pole Telescope team \citep{2010ApJ...718..632H} at 1.3 and 2 mm. More recently \citet{2011arXiv1105.1463P} measured the clustering signal by removing accurately the cirrus contamination at 100 and 160~$\mu$m. The power spectrum of the CIBA has also been computed using Herschel/SPIRE at 250, 350 and 500~$\mu$m \citep{2011arXiv1101.1080A}, taking advantage of its angular resolution, and using Planck/HFI at 350, 550, 850~$\mu$m and 1.3 mm \citep{2011arXiv1101.2028P} taking advantage of its sky coverage. Therefore, the clustering of infrared star-forming galaxies in the CIBA has been detected over a large range of wavelengths and angular scales. All these results have been analyzed in several ways, hardly comparable. As a first analysis, \citet{2007ApJ...665L..89L} derived the linear bias, the proportionality coefficient between the fluctuations of the dark matter (DM) density field and emissivities of galaxies. They found $b=2.4\pm0.2$ and \cite{2009ApJ...707.1766V} found $b=3\pm0.3$ which implies that these galaxies are a highly biased tracer of DM. The difference between these two biases may be explained by the fact that at longer wavelength, higher redshift infrared star-forming galaxies are probed \citep{2005ARA&A..43..727L,2008A&A...481..885F} and thus are found to be more biased. New measurements needed more complex models. \citet{2009ApJ...707.1766V}, \citet{2011arXiv1101.1080A} and \citet{2011arXiv1101.2028P} introduced a halo occupation distribution for the study of CIBA. It describes the DM distribution and especially how galaxies are distributed in one DM halo. It appears that each wavelength must be fitted separately which indicates an evolution of the clustering with the redshift \citep{2011arXiv1101.2028P}. Most of the models determined the mass of the halos where infrared star-forming galaxies lie and thus where star formation occurs. \\
In the long term purpose of analysing all these new measurements in a consistent way, we present a new model of the clustering in CIBA. We use the halo model formalism (\cite{2002PhR...372....1C}) which has been often used in the last few years to predict and to interpret galaxy clustering. We link it to a recent model of infrared star-forming galaxies evolution that reproduces well number counts and luminosity functions \citep{2011A&A...529A...4B}. This model of clustering has been successfully used to fit Planck data \citep{2011arXiv1101.2028P}.\\
The paper is organized as follows. We describe the model and its parameters in Sect. 2. We set a fiducial model inspired from \citet{2009ApJ...707.1766V,2011arXiv1101.2028P,2011arXiv1101.1080A} and compute angular power spectra for several instruments with which we carry out a Fisher analysis in Sect. 4. Sect. 5 is dedicated to interpreting measurements such as the redshift and halo-masses contribution to the power spectrum, the linear bias, the influence of the mean emissivities and the contribution of LIRGs and ULIRGs to power spectra. We finally conclude in Sect. 6. Throughout this study we use the Wilkinson Microwave Anisotropy Probe 7-year Cosmology \citep{2011ApJS..192...16L}.  
%---------------------------------------------
\section{Why a new model?}\label{par:why}
%---------------------------------------------
As said previously several models of clustering in the Cosmic Infrared Background already exist so why constructing a new one? There have been several measurements of the clustering and different models have been applied to analyze each measurement. Moreover their approaches are different which make comparison of the results difficult if not impossible. Thus one single model that ties together all available measurements is appealing, especially to analyze them simultaneously and search for an evolution of the clustering. Such a model requires three ingredients: a DM distribution, a relation between galaxies and DM halos and an evolution of infrared star-forming galaxies.\\
\citet{2007ApJ...665L..89L}, \citet{2007ApJ...670..903A} and \citet{2009ApJ...707.1766V} used the model of galaxy evolution of \citet{2003MNRAS.338..555L}. This model was the most up to date model at that time. It fitted well differential number counts and luminosity functions measurements from 24 to 850~$\mu$m (IRAS, Spitzer/SCUBA). However it does not reproduce very well new measurements, especially differential number counts from Herschel. This is a phenomenological model in which the evolution of the luminosity function was tuned to reproduce the constraints available at that time. It over-predicts the luminosity density at high-z. Moreover, it does not reproduce very well the observed redshift distribution of the CIB \citep{2011A&A...525A..52J}. It predicts a peak at $z\sim1$ that is not observed. The angular power spectra of CIBA strongly depend on the redshift distribution of the sources through the emissivities (see Sect. \ref{par:angular_pk}). Therefore, a `valid' distribution in redshift is important and a more robust model in agreement with most recent measurements is needed. \\
\citet{2010ApJ...718..632H} used the galaxy templates from \citet{2003MNRAS.338..555L} in order to check a simple model with a `single SED'. This model has only a few parameters that can be changed easily and thus adapted to each of their wavelengths. They fixed the shape of the power spectrum and only changed its amplitude depending on the wavelength. \citet{2007ApJ...670..903A} used the same model of infrared galaxies evolution : they used the luminosity function as a function of redshift at 350~$\mu$m coming from \citet{2003MNRAS.338..555L} that they matched to conditional luminosity functions (CLFs). Other wavelengths are extrapolated from the 350~$\mu$m. Finally Amblard et al. 2011 avoid using any model of galaxies evolution by letting free the redshift distribution of the cumulative flux coming from the background faint galaxies in several redshift bins. \\
Concerning the distribution of DM, \citet{2009ApJ...707.1766V} and \citet{2011arXiv1101.1080A} used the formalism of the halo model and the same halo occupation number whereas \citet{2007ApJ...665L..89L} and \citet{2010ApJ...718..632H} considered a linear power spectrum for dark matter. \citet{2007ApJ...670..903A} also used the halo model formalism through CLFs. By integrating CLFs on the luminosity, the halo mass function is recovered. However this approach depends on too many parameters that cannot be constrained simultaneously. \\
In order to construct a new model, we link an up to date model of galaxies evolution to a recent version of the halo model. We use the model of evolution of galaxies of \citet{2011A&A...529A...4B}. It reproduces well Herschel measurements as well as older ones (from 15 \um~to 1.1 mm). It also very well reproduces the redshift distribution of the CIB of \citet{2011A&A...525A..52J}. We use an updated version of the halo model of \citet{2009ApJ...707.1766V}, the halo occupation distribution (HOD) introduced by \citet{2010ApJ...719...88T}. This HOD reproduces well the angular correlation function of optical galaxies, red (star-forming) and blue (quiescent) galaxies at $0.4<z<2$. Therefore we make a strong assumption here, assuming that this description would work on star-forming galaxies. Given the current lack of understanding of the details of the process of star-formation and its evolution with redshift, it is difficult to define what would be a better HOD prescription and we therefore stay with this one. We study power spectra coming from our model for several wavelengths/instruments: 100~$\mu$m IRAS, 160~$\mu$m Spitzer/MIPS, 250, 350, 500~$\mu$m Herschel/SPIRE and 850~\um, 1.3 and 2mm Planck/HFI. A list of the available data of CIBA power spectra is given in table \ref{tab:shot_noise}.
%----------------------------------------------------------------------------------------
\section{The model}\label{par:mod}
%----------------------------------------------------------------------------------------
%----------------------------------------------------------------------------------------
\subsection{The parametric model of star-forming galaxies evolution}\label{par: gal_mod}
%----------------------------------------------------------------------------------------
To reproduce the angular power spectrum of the CIBA we need a model for the redshift evolution of star-forming galaxies. We use the model presented in \citet{2011A&A...529A...4B}. It is a backward evolution model based on a parametrized luminosity function and on galaxies spectral energy distribution templates.\\
\citet{2011A&A...529A...4B} consider a luminosity function (LF) that behaves like a power law for $L<<L^\star$ and like a Gaussian for  $L>>L^\star$ \citep{1990MNRAS.242..318S}~: 
\bea
\Phi(L_{IR})&=& \frac{dN(L_{IR})}{dVd\log_{10}(L_{IR})}\\
       &=&\Phi^{\star}(z)\left(\frac{L_{IR}}{L^{\star}(z)}\right)^{1-\alpha}\exp{\left[-\frac{1}{2\sigma^2}\log_{10}^2\left(1+\frac{L_{IR}}{L^{\star}(z)}\right)\right]}
\label{eq:lum_func}
\eea
where $\Phi(L_{IR})$ is the number of galaxies with the infrared bolometric luminosity $L_{IR}$ within the comoving volume $dV$ and the bin $d\log_{10}L$. $\Phi^{\star}$ is a normalization constant that fixes the density of sources. The low and high luminosity parts have different slopes, $1-\alpha$ and $1-\alpha-1/\sigma^2/\ln^2(10)$ respectively. $L^{\star}$ represents the luminosity at the break. The parameters that describe the luminosity function are listed in table \ref{tab:recap_param}.\\
The luminosity function evolves with the redshift through $L^{\star}$ and $\phi^{\star}$~: 
\bea
L^{\star}(z) &=& L^{\star}(z=0)(1+z)^{r_L}\\
\phi^{\star}(z) &=& \phi^{\star}(z=0)(1+z)^{r_{\phi}}
\eea
Exponents $r_L$ and $r_{\phi}$ are not identical for all $z$. Two breaks are imposed to reproduce the evolution of the LF. The first one $z_{break}$ is a free parameter and is found to be around 1. The second one is fixed at $z=2$ to avoid divergence at high $z$. Between these two breaks, the values of $r_L$ and $r_{\phi}$ change as shown on table \ref{tab:recap_param}.\\

\citet{2011A&A...529A...4B} used the SED library of \citet{2004ApJS..154..112L}. It contains two galaxy populations: star-forming and late-type galaxies. The latter emit half or less of their energy in the IR whereas the former emit more than 95 \% of their energy in the IR. The fraction of each population depends on luminosity. Indeed, late-type dominate at low luminosity whereas star-forming dominate at high luminosity. For a given bolometric luminosity, the fraction of star-forming is~: 
\bea
f_{SF}&=&\frac{\Phi_{SF}}{\Phi}\\
      &=& \frac{1+\tanh[\sigma_{pop}(L)\log_{10}(L_{IR}/L_{pop})]}{2}
\eea 
$L_{pop}$ is the luminosity where $\Phi_{SF}=\Phi_{late-type}$ and $\sigma_{pop}$ characterizes the width of the transition between the two populations.\\
Differential number counts are then derived for each population and then summed. At flux $S$~:  
\be
\frac{dN}{dS}(S) = \int_z\int_{L}f_{pop}\frac{dN_{pop}}{d\log_{10}L_{IR}dV}\frac{d\log_{10}L_{IR}}{dS}\frac{dV}{dz}dz
\ee
where dN/dS is the number of sources per flux unit in a unit solid angle and $pop$ = late-type or $pop$ = star-forming.\\
  \begin{figure}[!h]\centering 
    \includegraphics[scale=0.6]{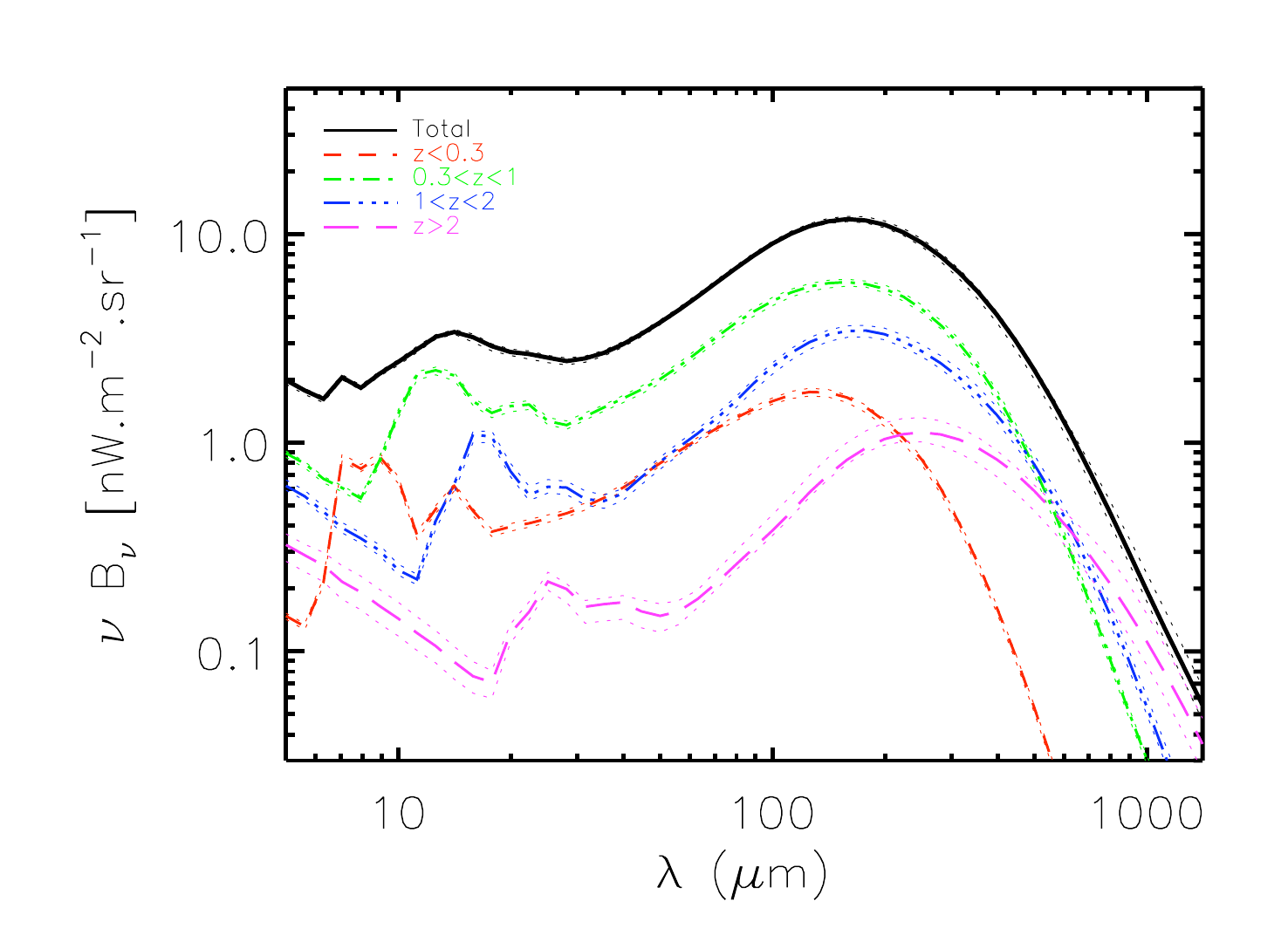}
    \caption{CIB per redshift bins from the model of \citet{2011A&A...529A...4B}. The high redshift contribution increases with the wavelength.}
    \label{fig:distrib_z_model_matt}
  \end{figure} 
The model of \citet{2011A&A...529A...4B} is described by thirteen free parameters. Best fit parameters and confidence areas are computed using Monte-Carlo Markov Chains on available and trustworthy differential number counts and luminosity functions at certain wavelengths. \citet{2011A&A...529A...4B} fitted number counts of Spitzer/MIPS at 24, 70 and 160~$\mu$m \citep{2010A&A...512A..78B}, those of Herschel/SPIRE at 250, 350 and 500~$\mu$m \citep{2010AA...518L..21O} and those of Aztec at 1.1 mm \citep{2010MNRAS.401..160A,2010MNRAS.405.2260S}. A couple of luminosity functions at different redshifts are also fitted, the 8~$\mu$m one at $z=2$ from \citet{2007ApJ...660...97C}, others derived from Rodighiero et al. 2009 (a local LF at 24~$\mu$m, a 15~$\mu$m one at $z=0.6$ and a 12~$\mu$m LF at $z=1$) and that at 60~$\mu$m at $z=0$ from \citet{1990MNRAS.242..318S}. Moreover, absolute measurements of the CIB are also used as a model constraint \citep{1999A&A...344..322L}. We do not describe and discuss all the fits here \citep[for a full discussion see][]{2011A&A...529A...4B}. Using the best fit, this model also provides the redshift distribution of the CIB as shown on Fig. \ref{fig:distrib_z_model_matt}. We see that higher-redshifts contribution increases with wavelength~: the redshift slice $0.3<z<1$ dominates up to 400~$\mu$m whereas in the sub-millimeter $z>2$ dominates. This model provides a very good agreement with the CIB redshift distribution \citep{2011A&A...525A..52J}. \\
We will study how these thirteen free parameters can be constrained with power spectra of the CIBA.

%------------------------------------------------------
\subsection{The angular power spectrum}\label{par:angular_pk}
\begin{figure}[!h]
\includegraphics[scale=0.5]{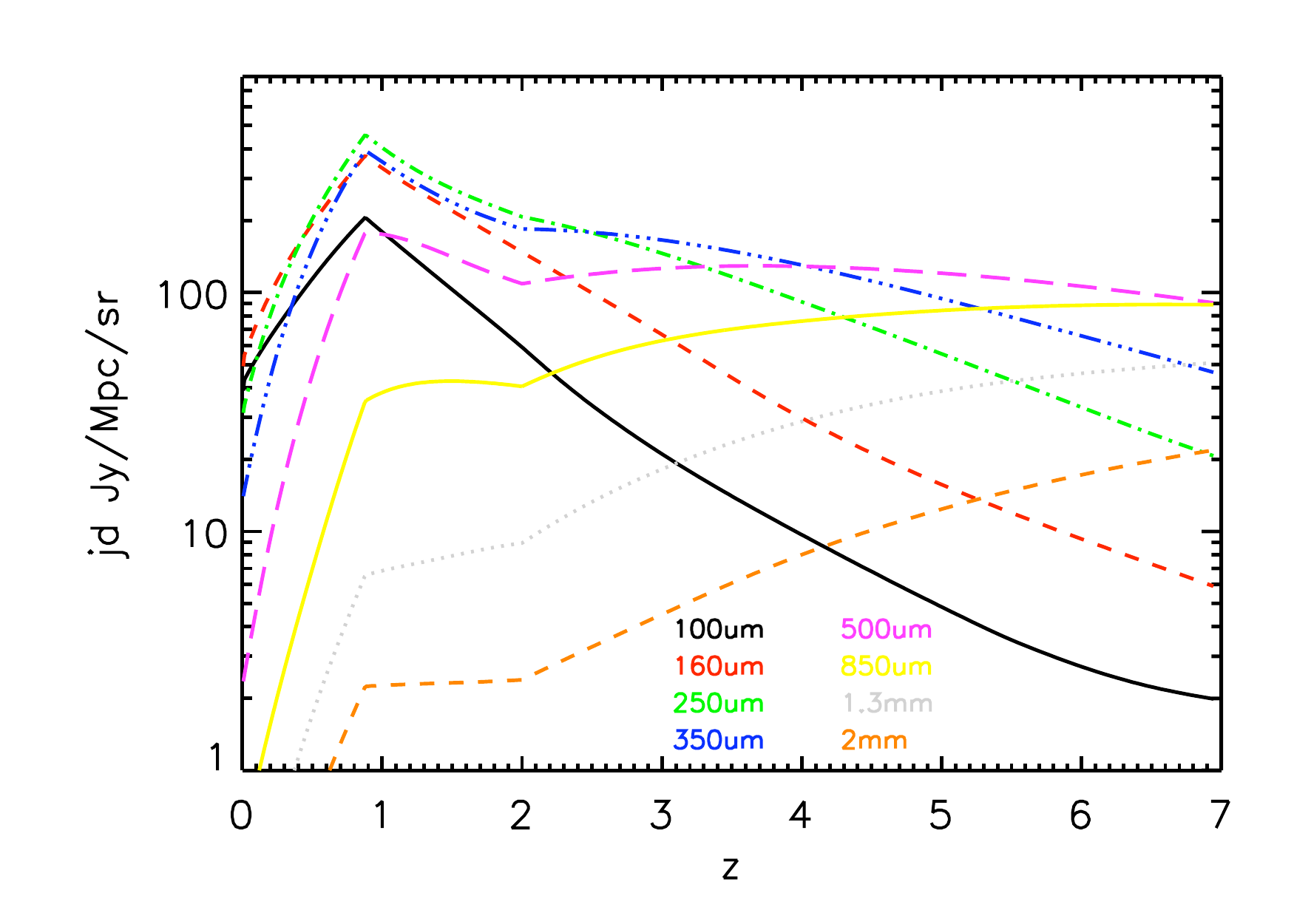}
\caption{Emissivities versus the redshift for different wavelengths. The contribution from high redshift increases with the wavelength.}
\label{fig:plot_jd}
\end{figure}

\begin{figure}[!h]
\includegraphics[width = \columnwidth]{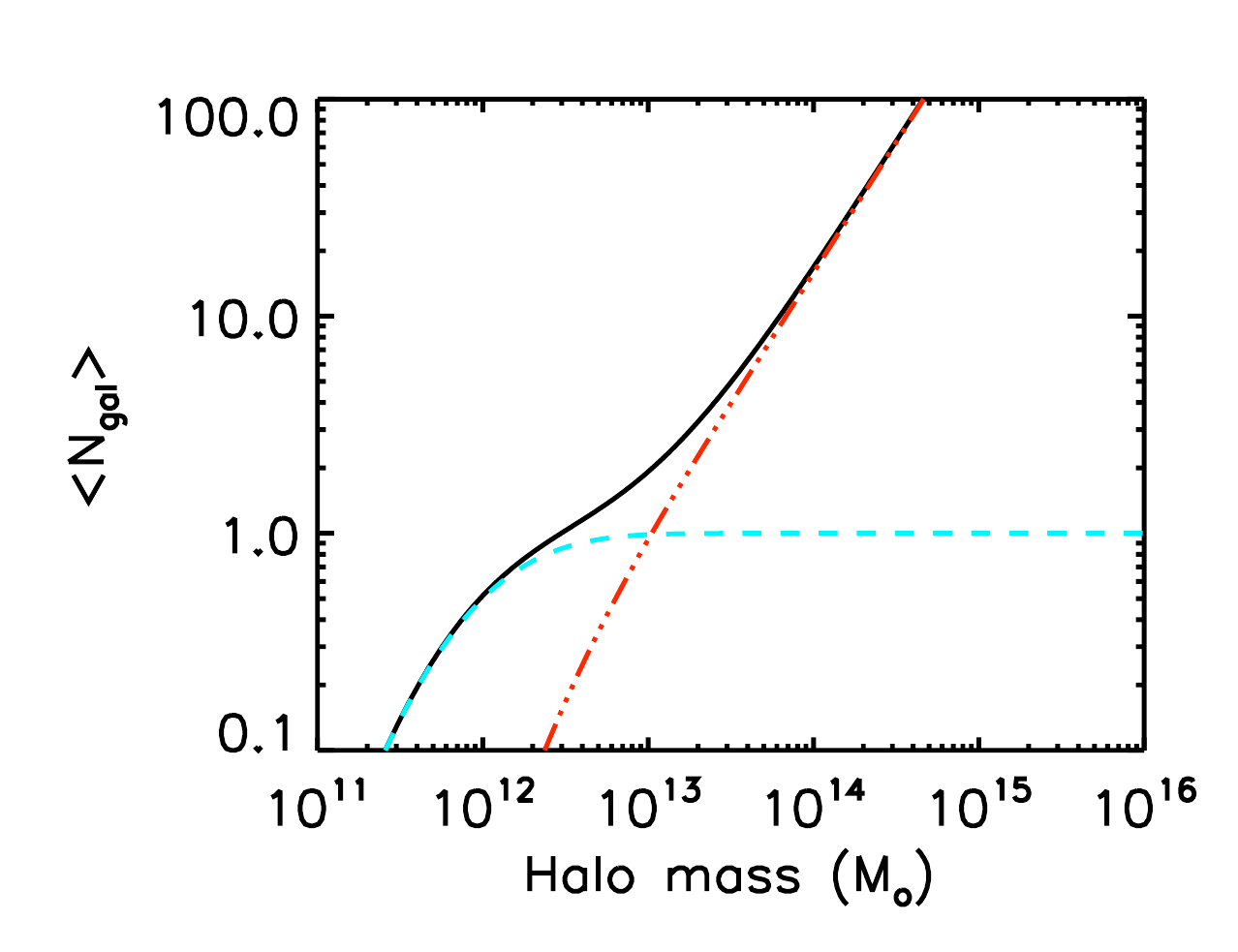}
\caption{Number of galaxies versus dark matter halo mass. The blue dashed line shows the central galaxies, the red dotted-dashed line shows satellite galaxies and the black continuous line shows the total. We use the parameters of our fiducial model (see Sect. \ref{par:pk+deg}), that is to say $\log M_{min}=11.5$, $M_{sat}=10M_{min}$ and $\alpha=1.4$}
\label{fig:gal_in_halo}
\end{figure}

%------------------------------------------------------
According to \citet{2000ApJ...530..124H}, \citet{2001ApJ...550....7K} and using the Limber approximation, the angular power spectrum of the anisotropies of the CIB at wavelengths $\lambda$ and $\lambda'$ is~: 
\be
C_\ell^{\lambda\lambda'} = \int{dz\frac{dr}{dz}\frac{a^2(z)}{d_A^2}\bar{j}_{\lambda}(z)\bar{j}_{\lambda'}(z)P_{ss}(k=\frac{\ell}{d_A},z)}
\label{eq:cl}
\ee
where $\ell$ is the multipole, $r$ is the conformal distance from the observer, $a(z)$ the scale factor, $d_A$ the comoving angular diameter distance, and $\bar{j}_{\lambda}(z)$ the mean emissivity per comoving unit volume at wavelength $\lambda$ as a function of $z$. When $\lambda=\lambda'$ we recover the auto power spectrum. $P_{ss}(k)$ is the galaxy three dimensional power spectrum.\\
The emissivities are computed using the parametric luminosity functions following~:
\be
j_{\nu}(z) = \left(a\frac{d\chi}{dz}\right)^{-1}\int_{L}S(L_{IR},z)\frac{dN}{dzd(lnL_{IR})}d(lnL_{IR})
\ee
where $dN/dzd(lnL)$ is the number of galaxies per redshift bin $dz$ and per luminosity bin $d(lnL)$ and $S$ the flux. Each galaxy population (late-type and star-forming) emissivity are computed and summed to get the overall emissivity. Fig. \ref{fig:plot_jd} shows emissivities as a function of redshift. The two discontinuities at $z\sim0.9$ and $z = 2$ are due to the breaks imposed by the parametrization of the model of galaxies. It is clear that as the wavelength increases, the contribution from the high redshift part increases. Emissivities are color corrected according to their instrument and wavelengths to give \cl~in Jy$^2$/sr (for the photometric convention $\nu I_{\nu}$=cst).\\
In the context of the halo model, $P_{ss}(k)$ is the sum of the clustering in one single halo (1h) and in two different halos (2h):
 \be
P_{ss}(k) = P_{1h}(k) +P_{2h}(k)
 \ee
where 
\bea
P_{1h}(k) & = & \int_M {dN\over dM}{\langle N_{gal}(N_{gal}-1) \rangle \over \bar n_{gal}^2} U(k,M)^p dM \\
P_{2h}(k) & = & P_{lin}(k)\left[\int_M {dN\over dM}b(M){\langle N_{gal} \rangle \over \bar n_{gal}} U(k,M) dM\right]^2\\ .
\label{eq:p12h}
\eea
Here M is the halo mass, $P_{lin}(k)$ is the dark matter linear power spectrum \citep[computed with the fit of][]{1998ApJ...496..605E}, $U(k,M)$ the normalized Fourier transform of the halo density profile that is assumed to be that of \citet{1996ApJ...462..563N} truncated at the virial radius. $b(M)$ is the halo bias, $\langle N_{gal} \rangle$ the probability of having $N_{gal}$ galaxies in a halo of mass $M$ and we consider $p=2$ \citep{2002PhR...372....1C}. The mean number density of galaxies $\bar n_{gal}$ is given by:
\be
\bar{n}_{gal} = \int{\frac{dN}{dM}\langle N_{gal}\rangle dM}
\ee
where $dN/dM$ is the halo mass function. We will use the universal form given by \cite{2008ApJ...688..709T}  as well as its redshift evolution. We use its associated halo bias (see Eq. A1 in \citet{2009arXiv0902.1748T}). \\
The halo occupation number introduces galaxies in the halos statistically. Recent data and simulations suggest a necessary distinction between the major galaxy that lies at the center of the halo and the satellite galaxies that populate the rest of the halo. Above a given mass threshold, most halos will host a central galaxy. Above a second higher mass threshold, they will also host satellite galaxies. $N_{gal}$ can thus be written as : 
\be
\langle N_{gal}\rangle  =\langle N_{cen}\rangle+\langle N_{sat}\rangle\ .
\ee
According to the prescription of \cite{2010ApJ...719...88T}, the occupation function of central galaxies is: 
\be
\langle N_{cen}\rangle = \frac{1}{2}\left[1+\mbox{erf}\left(\frac{\log M-\log M_{min}}{\sigma_{\log M}}\right)\right]\label{eq:hod_cent}
\ee
where $M_{min}$ is the halo mass at which a halo has a 50 \% probability of hosting a central galaxy. $\sigma_{\log M}$ controls the width of the transition between zero and one central galaxy. There is a smooth transition between low mass halos that do not contain bright enough galaxies to be seen in the data ($M<<M_{min}$) and more massive ones that always contain a bright central galaxy. ($M>>M_{min}$). The satellite occupation function is: 
\be
\langle N_{sat}\rangle  = \frac{1}{2}\left[1+\mbox{erf}\left(\frac{\log M-\log 2M_{min}}{\sigma_{\log M}}\right)\right]\left(\frac{M}{M_{sat}}\right)^{\alpha_{sat}}\label{eq:hod_sat}
\ee
It has a cut-off of the same form as the central occupation with a transition mass twice larger than that of the central to prevent halos which have a low probability of hosting a central galaxy to contain satellite galaxies. The number of satellite galaxies grows with a slope of $\alpha_{sat}$. Both number of galaxies as well as their sum are plotted on Fig. \ref{fig:gal_in_halo}. \\
With this model, the angular power spectrum of CIBA depends on only four halo model parameters $\alpha_{sat}$, $M_{min}$, $M_{sat}$ and $\sigma_{logM}$. Cosmology is fixed at WMAP7 values. Our parameters are listed in Table \ref{tab:recap_param} with their meaning and their fiducial values that we set in Sect. \ref{par:pk+deg} . \\
The long term purpose of our model is to look for best fits of these parameters for Spitzer/MIPS, IRIS, Planck, Herschel and SPT data and study their evolution with wavelength. However it is beyond the scope of this paper. \ttb{Therefore we will not compare the data to the power spectra coming from our model}. Our first aim here is to study the parameter space and to investigate particularly the behaviour of the halo bias, the halo mass-contribution to the power spectrum and its redshift distribution. To do so we will consider a set of fiducial halo parameters identical at all wavelengths.

\begin{table*}\centering
 \begin{tabular}{{lll}}
 \hline\hline
 parameter name	        & Description                                                            & Value\\
\hline
$M_{min}$		& Minimal mass of a halo to have a central galaxy                        & 10$^{11.5}M_{\odot}$\\
$M_{sat}$		& Nomalisation mass for satellite galaxies                               & 10$^{12.5}M_{\odot}$\\
$\alpha_{sat}$		& Slope of the number of satellite galaxies at high mass                 & 1.4\\
$\sigma_{logM}$          & Scatter in halo mass                                                   & 0.748\\
 \hline
$\alpha$        	& Faint end slope of the IR bolometric LF                                &1.223\\
$\sigma$		& Parameter driving the bright end slope                                 &0.406\\
$L_{\star}$(z=0) 	& Local characteristic luminosity of the LF                              &2.377$\times10^{10}L_{\odot}$\\
$\phi_{\star}$(z=0)	& Local characteristic density of the LF                                 &3.234$\times10^{-3}$gal/dex/Mpc$^3$\\
$r_{L_{\star},lz}$ 	& Evolution of the characteristic luminosity between 0 and $z_{break}$    &2.931\\
$r_{\phi_{\star},lz}$ 	& Evolution of the characteristic density between 0 and $z_{break}$       &0.774\\
$z_{break}$		& Redshift of the first break                                            &0.879\\
$r_{L_{\star},mz}$ 	& Evolution of the characteristic luminosity between $z_{break}$ and 2    &4.737\\
$r_{\phi_{\star},mz}$ 	& Evolution of the characteristic density between $z_{break}$ and 2       &-6.246\\
$r_{L_{\star},hz}$ 	& Evolution of the characteristic luminosity for $z>2$                   &0.145\\
$r_{\phi_{\star},hz}$ 	& Evolution of the characteristic density for $z>2$                      &-0.919\\
$L_{pop}$		& Luminosity of the transition between normal and star-formig templates  &23.677$\times10^{10}$\\
$\sigma_{pop}$  	& Width of the transition between normal and star-forming templates      &0.57\\
\hline
\end{tabular}
\caption{Parameters of our model. The first part of the table lists the halo model parameters and the second part lists the parameters of the model of galaxies. The values of the latter are the mean ones of \citet{2011A&A...529A...4B}.}
\label{tab:recap_param}
\end{table*}

%-------------------------------------------------------------
\section{Power spectra and parameters degeneracies}\label{par:pk+deg}
%-------------------------------------------------------------

In this section, we present the CIB power spectra computed with the model detailed in the previous section for several wavelengths in the far-IR and sumillimeter. We then study the degeneracies of the parameters, looking first at the galaxies model parameters and second at the HOD parameters.

\subsection{Power spectra}

\begin{figure*}[!ph]
    \includegraphics[width=\textwidth]{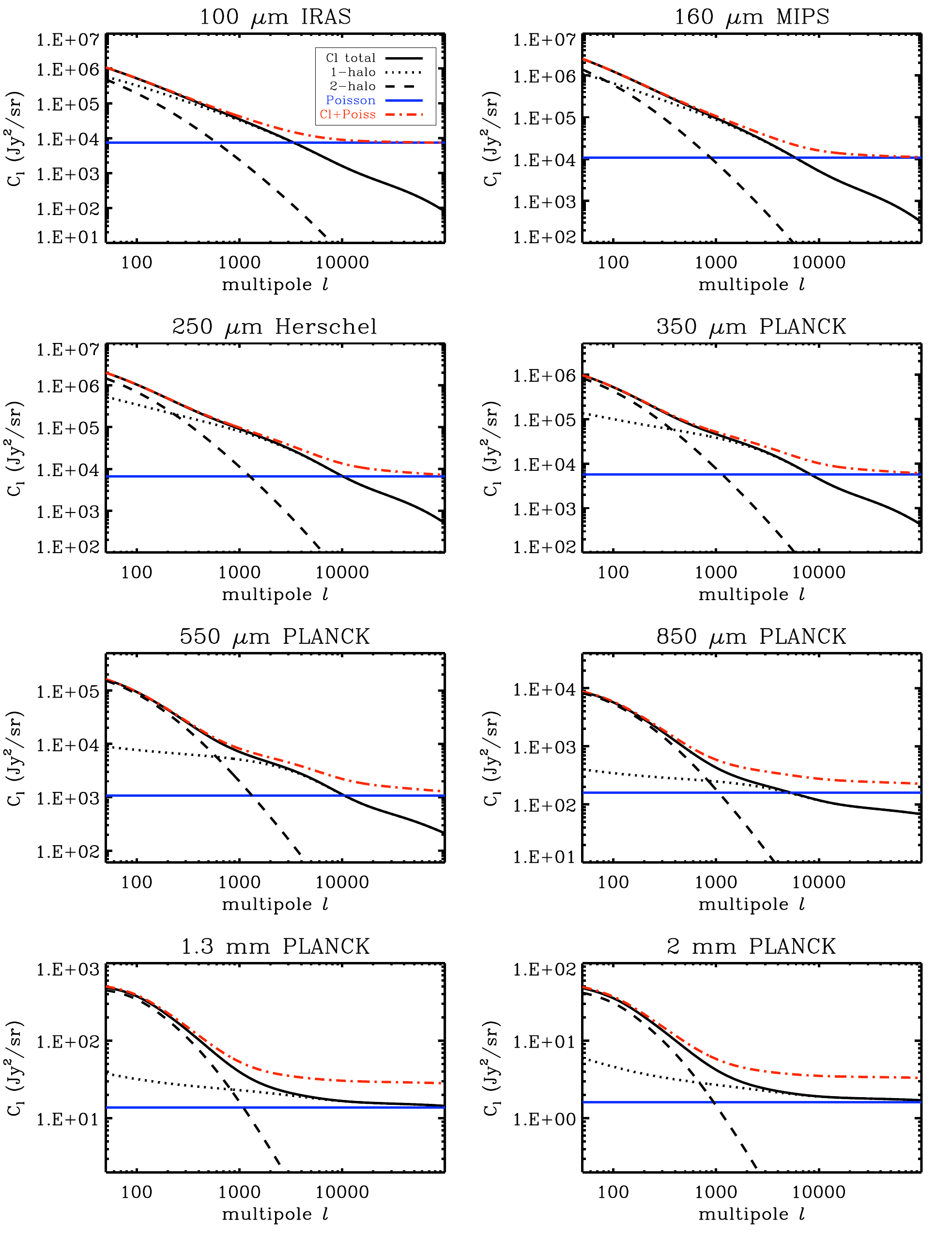}
%scale=0.85]{cl_tot+pois}
    \caption{CIB anisotropies power spectra at several wavelengths. The continous black line shows the power spectra of the clustering. The dotted black line is the 1-halo term of the power spectrum and the dashed line is the 2-halo term. The blue horizontal line represents the shot noise level and the red dot-dashed curve is the total power spectrum. Titles of the plots give the reference for the shot noise level and the used bandpass filters.}
    \label{fig:cl_tot+pois}
  \end{figure*}

Our fiducial model is set with the HOD parameters $\log M_{min}=11.5$, $M_{sat}=10M_{min}$ and $\alpha=1.4$ at all wavelengths. These values are motivated by the parameters fit of \citet{2009ApJ...707.1766V}, \citet{2011arXiv1101.2028P}, \citet{2011arXiv1101.1080A}. As the halo parameters slightly depend on the wavelength (in reality but it is not the case here), the power spectra presented in this section may not be seen as an exact prediction but as a basis for a qualitative study. For this fiducial model, we present on Fig. \ref{fig:cl_tot+pois} the power spectra for different experiments and selected wavelengths, from 100~$\mu$m to 2 mm. \\

The comparison to measurements also requires the introduction of a shot noise term due to the finite number of galaxies. We compute it using our galaxy evolution model \citep{2011A&A...529A...4B}.
\be 
C_{\ell} = C_{\ell,clus}+C_{\ell,shot}
\ee
where $C_{\ell,clus}$ is the power spectrum of the clustering and $C_{\ell,shot}$ is the shot noise.  $C_{\ell,shot}$ depends on the flux cut applied to the data when removing or masking the brightest sources. Typical flux cuts for different far-IR and sub-millimeter experiments are given in Tab. \ref{tab:shot_noise}. On Fig. \ref{fig:cl_tot+pois}, we only show one shot noise level per wavelength corresponding to the instrument given in the plot titles, for purpose of clarity.\\
The contribution of the 1-halo (2-halo) corresponds to the dotted line (dashed line). The instrument noise is not shown here but it is often negligible, the CIB being measured with a very high S/N even at spatial scales close to the angular resolution. Looking at this plot, we see clearly that the interplay between 2h, 1h and shot noise terms will make the interpretation of measurements quite subtle. The contribution of the 1-halo term decreases with the wavelength which can prevent its measurement if the resolution of the instrument is low. For example, Planck with its 5 \arcmin angular resolution at high frequency, cannot measure directly the shot noise level and the 1-halo term can be easily misinterpreted as shot noise. Reversely, the 1-halo term dominates a large range of scales at 100 and 160~$\mu$m and thus can be measured accurately at these wavelengths. \\
High wavelengths probe higher redshifts than short ones and halos are relatively smaller at high redshifts compared to those in the local Universe as the latter had time to accrete more matter. Therefore the scale of the intersection between the 1- and 2-halo terms shifts towards higher $\ell$ as the wavelength increases. 
It goes from $\ell\sim50$ at 100~$\mu$m to $\ell\sim1000$ at 2~mm. \cite{2009ApJ...707.1766V} also observed this trend. At 250~$\mu$m the crossing is at $k\sim0.03$ arcmin$^{-1}$ ($\ell\sim648$) whereas it is at $k\sim0.06$~arcmin$^{-1}$ ($\ell\sim1296$) at 500~$\mu$m. The exact crossing point differs from ours because of the HOD parameterization.

\begin{table*}\centering
 \begin{tabular}{*{5}{c}}
 \hline\hline
 wavelength (\um)    &Instrument            &Reference                     & Flux Cut (mJy)      & Shot noise level (Jy$^2$/sr)\\
\hline
100                  &IRIS                  & \citet{2011arXiv1105.1463P}  & 700                & 7364$\pm$1232\\
\hline
160                  &Spitzer/MIPS          & \citet{2007ApJ...665L..89L}  & 200                & 10834$\pm$3124\\
\hline
250                  &Herschel/SPIRE        & \citet{2011arXiv1101.1080A}  & 50                 & 6715$\pm$1458\\
\hline
350                  &Herschel/SPIRE        & \citet{2011arXiv1101.1080A}  & 50                 & 4362$\pm$1250\\
350                  &Planck/HFI            & \citet{2011arXiv1101.2028P}  & 710                & 5923$\pm$367\\
\hline
500                  &Herschel/SPIRE        & \citet{2011arXiv1101.1080A}  & 50                 & 1156$\pm$434\\
550                  &Planck/HFI            & \citet{2011arXiv1101.2028P}  & 540                & 1150$\pm$92\\
\hline
850                  &Planck/HFI            & \citet{2011arXiv1101.2028P}  & 325                & 138$\pm$22\\
\hline
1363                 &SPT                   & \citet{2010ApJ...718..632H}  & 6.4                & 11.9$\pm$4.0\\
1363                 &ACT                   & \citet{2010AAS...21538407F}  & 20                 & 12.5$\pm$3.9\\
1380                 &Planck/HFI            & \citet{2011arXiv1101.2028P}  & 160                & 12.9$\pm$2.9\\
\hline
2000                 &SPT                   & \citet{2010ApJ...718..632H}  & 6.4                & 1.73$\pm$0.54\\
2000                 &ACT                   & \citet{2010AAS...21538407F}  & 20                 & 1.78$\pm$0.60\\
2097                 &Planck/HFI            &                              & 245                & 1.4$\pm$0.3\\
\hline
\end{tabular}
\caption{Shot noise levels in Jy$^2$/sr from \cite{2011A&A...529A...4B} for available measurements of CIBA power spectra.}
\label{tab:shot_noise}
\end{table*}

\subsection{Variation of power spectra with the galaxy-evolution model parameters}\label{par:gal_mod_param}

\begin{figure*}
\begin{tabular}{cc}
\includegraphics[scale=0.48]{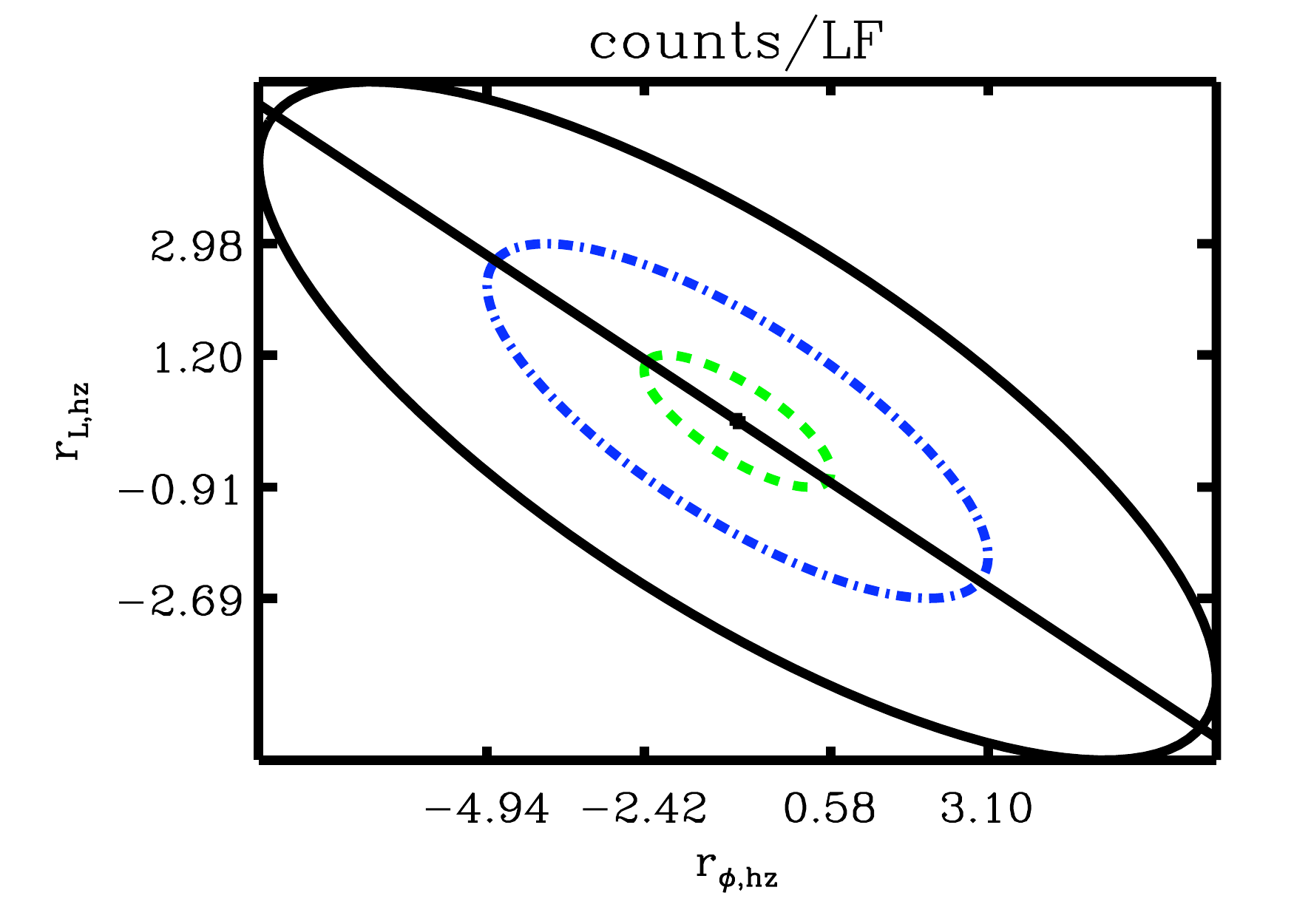}&
\includegraphics[scale=0.48]{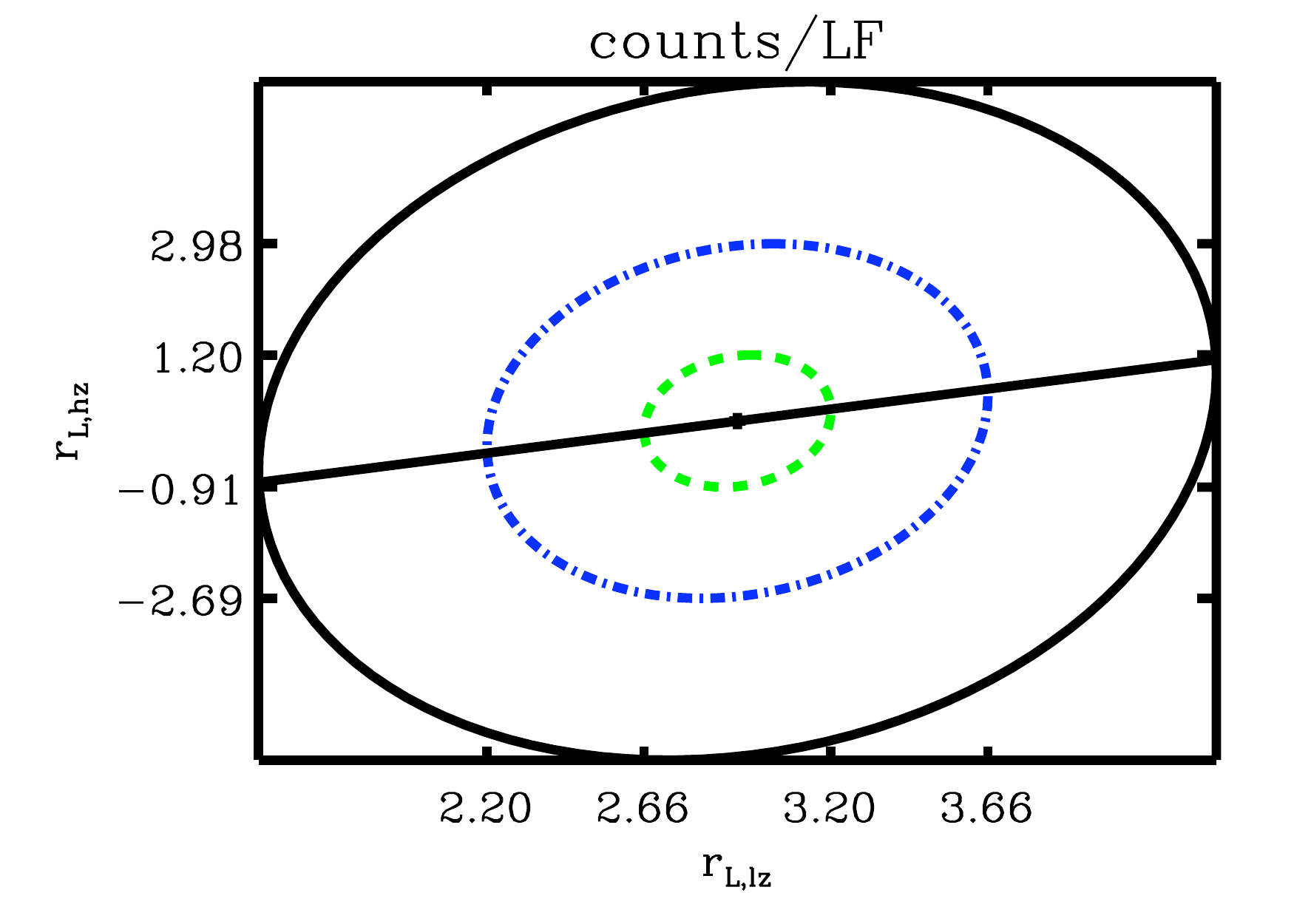}\\
\includegraphics[scale=0.48]{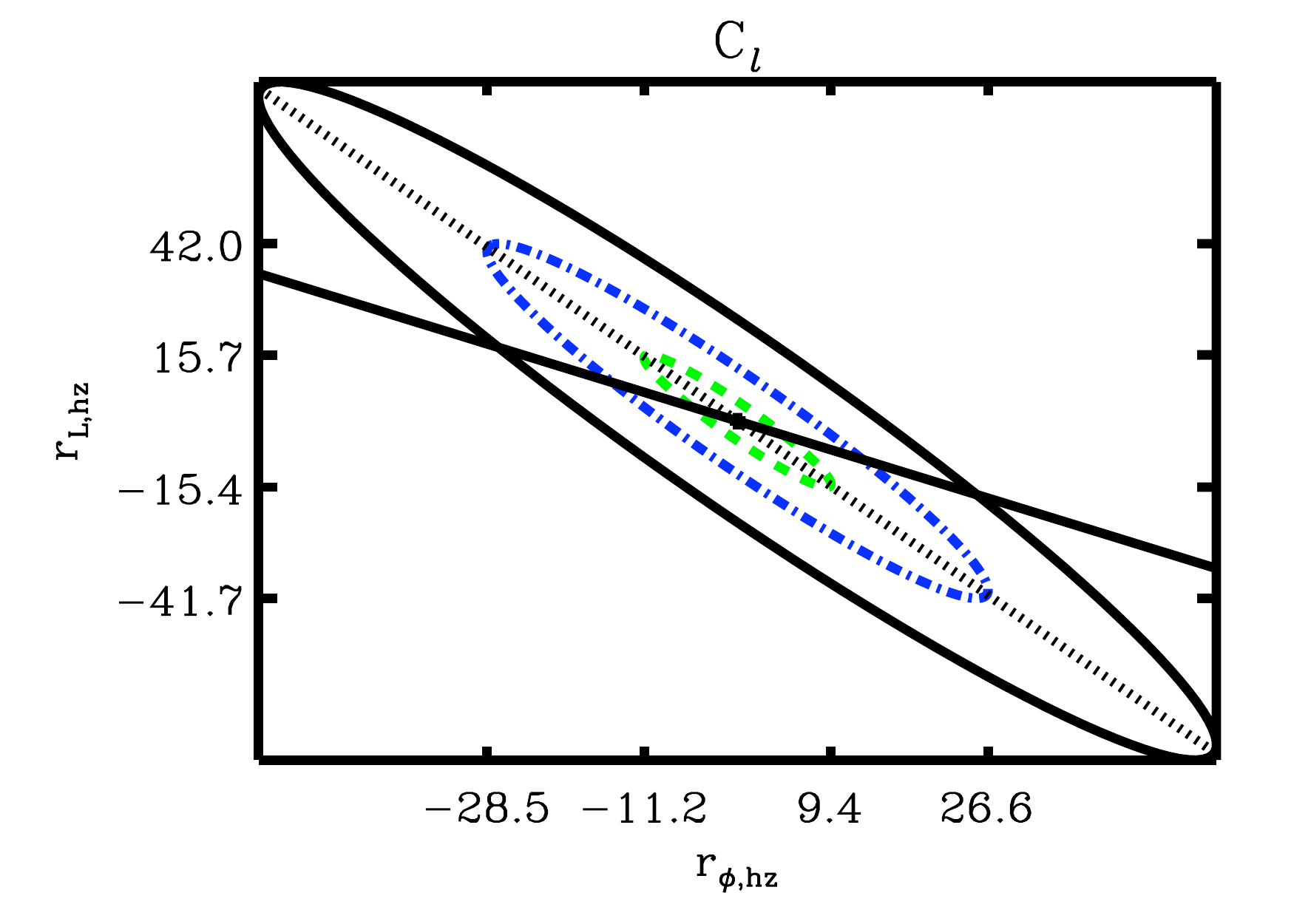} &
\includegraphics[scale=0.48]{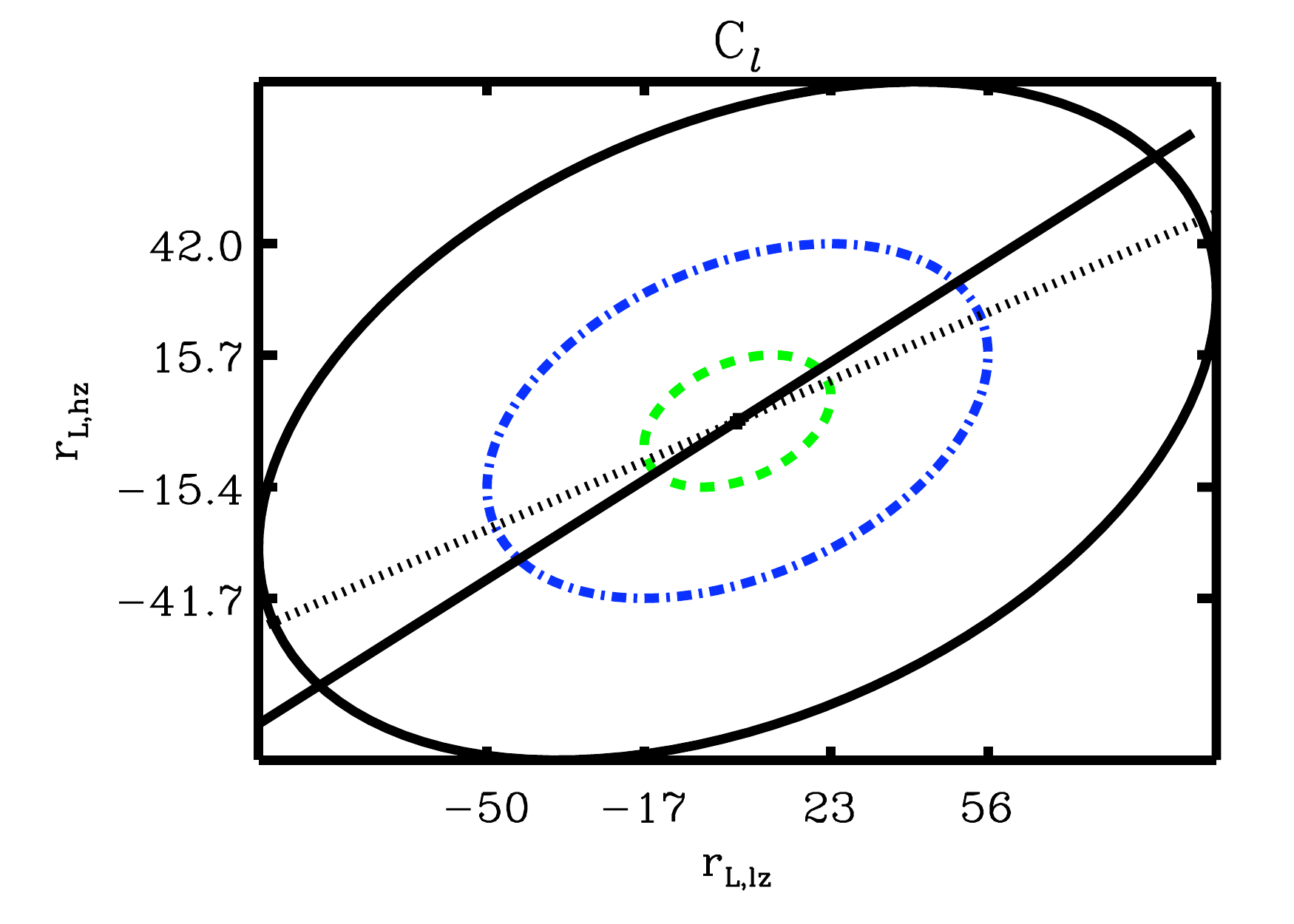} \\
\includegraphics[scale=0.48]{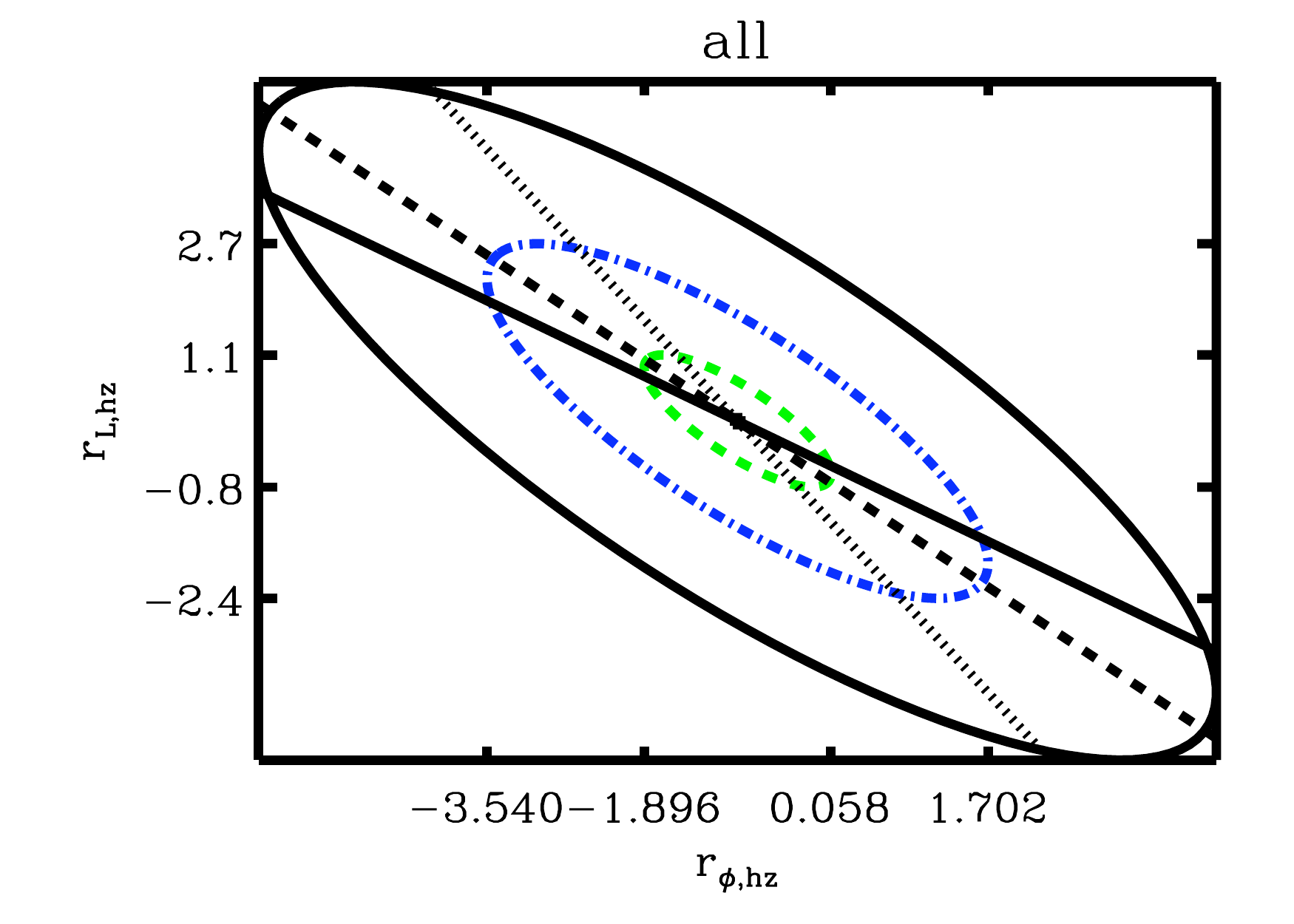}&
\includegraphics[scale=0.48]{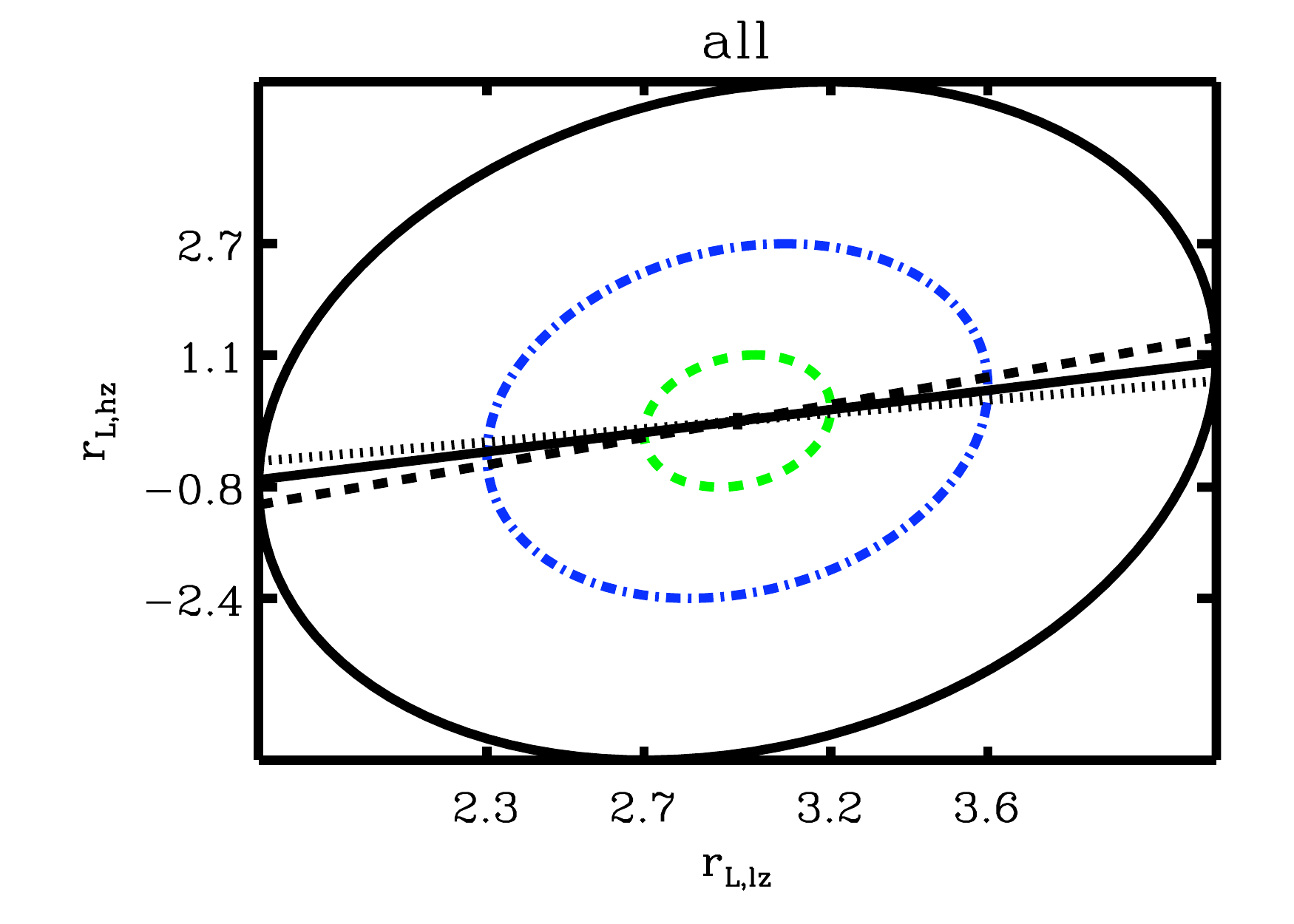}\\
\end{tabular}
\caption{ 1$\sigma$ (dashed green), 2$\sigma$ (dash-dot blue), 3$\sigma$ (black) likelihood contours of the galaxy evolution model parameters $r_{L^{\star},hz}$, $r_{L^{\star},lz}$ and  $r_{\phi^{\star},hz}$. Top, Middle and Bottom panels show the contours computed using counts/LF data, \cl~data only, and combined counts/LF and \cl~data, respectively. 
The continous line shows the direction of degeneracy using only counts/LF, the dotted line shows that using \cl~and the dashed line is that using all data.
$C_\ell$ alone cannot constrain the galaxy-evolution model parameters and the combination of \cl~and counts/LF data do not change much the constraints. It slightly improves them.}
 \label{fig:param_gal}
 \end{figure*}

In order to do an overall study of how our model parameters can be constrained, i.e. to investigate how degenerated they are, we construct the Fisher matrix associated to the power spectra. We write the Fisher matrix for angular power spectrum measurements as:
\be 
F_{ij} = \sum_{\lambda}\sum_\ell \frac{1}{\sigma_\ell^{\lambda 2}}\frac{\partial C_\ell^\lambda}{\partial \theta_i}\frac{\partial C_\ell^\lambda}{\partial \theta_j}\ 
\ee
where $\sigma_\ell$ are the errors on the measurements and they include both the cosmic variance and the instrumental noise at a multipole $\ell$:
\be
\sigma_{\ell}^2 = \left(C_\ell + \frac{N_\ell}{B_\ell^2}\right)^2\frac{2}{f_{sky}(2\ell+1)}
\label{Eq:eb}
\ee
 where $f_{sky}$ is the fraction of the sky we consider, $N_\ell$ is the level of the instrumental noise and $B_\ell^2$ the power spectrum of the beam. To compute the Fisher matrices, we generate mock power spectra using our fiducial model and error bars derived following Eq.~(\ref{Eq:eb}), from 100~$\mu$m to 1.3~mm. The range of multipoles is taken to be consistent with the available data. At 350 and 550~$\mu$m, we assume combined power spectra for Planck and Herschel, and we thus extend Planck power spectra to the Herschel limit in multipole. We plot \cl~in Jy$^2$/sr. They can be converted in $\mu$K$^2$ using the coefficients given in Tab. \ref{tab:conv}.\\
\begin{table}%[!t]\centering
 \begin{tabular}{*{3}{c}}
 \hline\hline
wavelength (\um)    & Instrument      & Jy$^2$/sr to $\mu$K$_{CMB}^2$\\
\hline
100           & IRAS            & 9.59 $\times10^{22}$\\
160           & Spitzer/MIPS    & 3.12 $\times10^{11}$\\
250           & Herschel/SPIRE  & 1.34 $\times10^{3}$\\
\hline
350           & Herschel/SPIRE  & 2.78 $\times10^{-1}$\\
350           & Planck/HFI      & 2.00 $\times10^{-1}$\\
\hline
500           & Herschel/SPIRE  & 7.45 $\times10^{-4}$\\
550           & Planck/HFI      & 2.94 $\times10^{-4}$\\
\hline
850           & Planck/HFI      & 1.20 $\times10^{-5}$\\
\hline
1380          & SPT             & 4.39 $\times10^{-6}$\\
1380          & Planck/HFI      & 4.32 $\times10^{-6}$\\
\hline
2000          & SPT             & 6.10 $\times10^{-6}$\\
2097          & Planck/HFI      & 7.31 $\times10^{-6}$\\
\hline
\end{tabular}
\caption{Conversion factors from Jy$^2$/sr to $\mu$K$_{CMB}$. One should multiply the power spectrum in Jy$^2$/sr (with the convention $\nu I_\nu=$cst) by the coefficient to get $\mu$K$_{CMB}^2$.}
\label{tab:conv}
\end{table}
The bottom left panel of Fig. \ref{fig:param_gal} shows confidence ellipses (1$\sigma$ in green and 2$\sigma$ in black) coming from \cl~when trying to measure only the galaxy model parameters $r_{L^{\star},hz}$ and $r_{L^{\star},lz}$. Clearly, they are very poorly constrained. For instance $r_{L^{\star},hz}$ = 0.145$\pm15.55$, or $r_{L^{\star},lz}$ = 2.93$\pm20.0$. For reference, the constraints obtained using current number counts are $r_{L^{\star},hz}$ = 0.145$\pm1.05$ and $r_{L^{\star},lz}$ = 2.93$\pm0.27$, as shown on the top left panel of Fig. \ref{fig:param_gal}. The fact is that these parameters enter in the expression of the \cl~through the emissivities which are integrated on all redshifts, they are thus hard to measure from clustering measurements alone. 

\begin{figure}[!h]
\includegraphics[height=0.9\textheight]{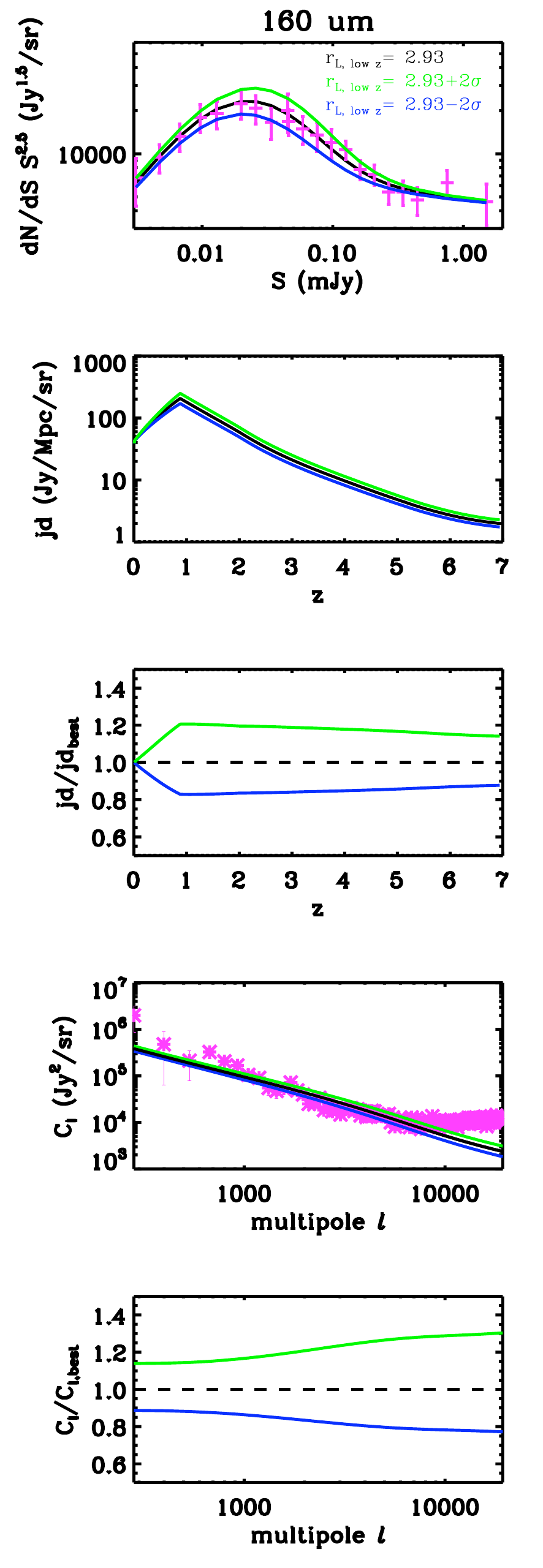}
\caption{Counts at 160~\um, emissivities and clustering power spectrum for three values of $r_{L_{\star},lz}$, the best fit and the best fit $\pm2\sigma$. Pink crosses are data, \citet{2010A&A...512A..78B} for the counts and \citet{2007ApJ...665L..89L} for the power spectrum.
Top panel : differential number counts at 160~\um.
2nd panel : mean emissivities at 160~\um.
3rd panel : ratio of the modified emissivities compared to the best fit one at 160~\um. 
4th panel : power spectra of the clustering at 160~\um.
Bottom panel : ratio of the modified power spectra compared to the best fit one at 160~\um.
A small change of $r_{L_{\star},lz}$ leads to a 20 \% modification on the emissivities and on 15-35\% on the \cl.}
\label{fig:var_lstar}
\end{figure}

As a matter of fact, the  lack of information in \cl~partially comes from the large number of parameters in the model of evolution of galaxies. To quickly quantify this we vary only a few of these parameters ($r_{L{\star},lz}$, $r_{L{\star},hz}$, $z_{break}$, $r_{\phi{\star},lz}$ and $r_{\phi{\star},hz}$) assuming that the others are perfectly known. Fixing all but these parameters corresponds to assuming that only the redshift evolution of the LF is unknown, clearly an irrealistic assumption. Not surprisingly, while some of the degeneracies remain in these reduced parameter space, on the whole, parameters are better contrained. For instance, we now obtain $r_{L^{\star},lz}$ = 2.93$\pm0.10$  and $r_{L^{\star},hz}$ = 0.145$\pm7.05$ which are about a two order of magnitude and a factor 2 improvement, respectively as compared to the numbers above.\\
In order to illustrate further this lack of information in \cl s, we show how they change with only one parameter, $r_{L^{\star},lz}$. We make it vary by $\pm2\sigma$ from its best fit \citep[$\sigma$ coming from][]{2011A&A...529A...4B}. This parameter governs the evolution of the luminosity function for $0<z<z_{break}$. A higher $r_{L^{\star},lz}$ means a faster increase of the luminosity, thus a higher value of $L^{\star}(z=z_{break})$. The top panel of Fig. \ref{fig:var_lstar} shows the influence of this parameter on the counts at 160~\um. A higher (smaller respectively) $r_{L^{\star},lz}$ implies higher (smaller) number counts thus more (less) galaxies on a large range of fluxes. This leads to a higher (smaller) emissivities as shows in the second and third panels of Fig. \ref{fig:var_lstar}. This results in a modification of $\sim20\%$ on the emissivities and from 15 to 35 \% on power spectra depending on the scale. The fact that this ratio is not constant is due to to the fact that the ratio of the emissivities is not constant with redshift (3rd panel of Fig. \ref{fig:var_lstar}). We can see that all power spectra are consistent within error bars and thus we can hardly discriminate between them. Therefore, it is hard to constrain the evolution model of galaxies using only power spectra. \\
More relevant data are required. We compare our confidence ellipses with those obtained with luminosity functions and number counts data. To do so we use the covariance matrix of \citet{2011A&A...529A...4B}. The error bars are in general much smaller and there are only a few degeneracies. For instance, as shown on the left panel of Fig. \ref{fig:param_gal}, $r_{L^{\star},hz}$ and $r_{\phi^{\star},hz}$ are still strongly degenerate but they are now much better constrained. \\
So far, we have investigated how galaxy evolution parameters are degenerated and constrained using LF/counts and \cl~separately. The next step is to look at the degeneracies when combining all these data. To do so, we add the two Fisher matrices coming from the counts/LF and the \cl. The bottom panels of Fig. \ref{fig:param_gal} show the confidence ellipses for $r_{L^{\star},hz}$, $r_{L^{\star},lz}$ and  $r_{\phi^{\star},hz}$ using the combined data. Note that the axis scales are different. The continous/dashed/dotted lines on Fig. \ref{fig:param_gal} indicate the degeneracy directions. They are different which illustrate the complementarity of the two data-sets and the constraints can be greatly improved. For example, the errors on $r_{\phi^{\star},hz}$ are decreased by a factor of 1.5 but the errors on $r_{L^{\star},hz}$ is not changing. However, this plot also clearly shows that overall, the number counts and LF measurements are much more powerful when looking at constraining the LF.
%This comes from sub 0.145$\pm0.95$ instead of $r_{L^{\star},hz}$ = 0.145$\pm1.05$ with the counts/LF. The same goes for $r_{L^{\star},lz}$, $r_{\phi^{\star},hz}$ and the other parameters of the model, the constraints stay unchanged or are slightly improved. We can then hope to constrain better the model of evolution of galaxies by using the CIBA power spectra}.\\ 
 However, \cl~can still constrain the global evolution of galaxies through their mean emissivities. A first attempt was made by \citet{2011arXiv1101.1080A} who did not use a model of galaxies to compute the emissivities but bin them in several redshift intervals and considered the values of the emissivities in these four bins as free parameters. They also required that the integrated source density is within the 68\% confidence level ranges of the CIB obtained by FIRAS. 

\subsection{Halo occupation distribution parameters and their degeneracies}
\begin{figure*}[!t]
\includegraphics[scale=0.35]{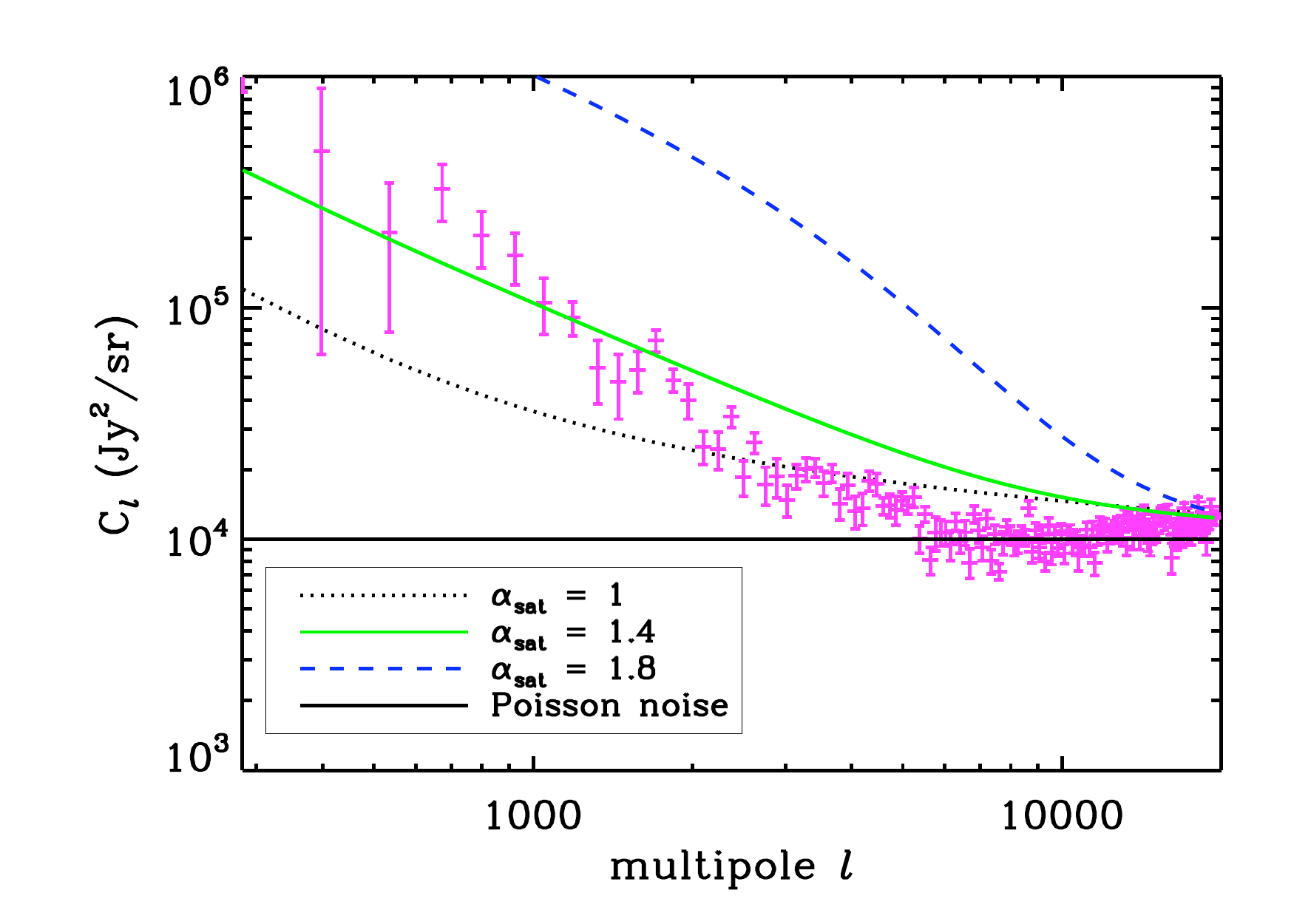}\includegraphics[scale=0.35]{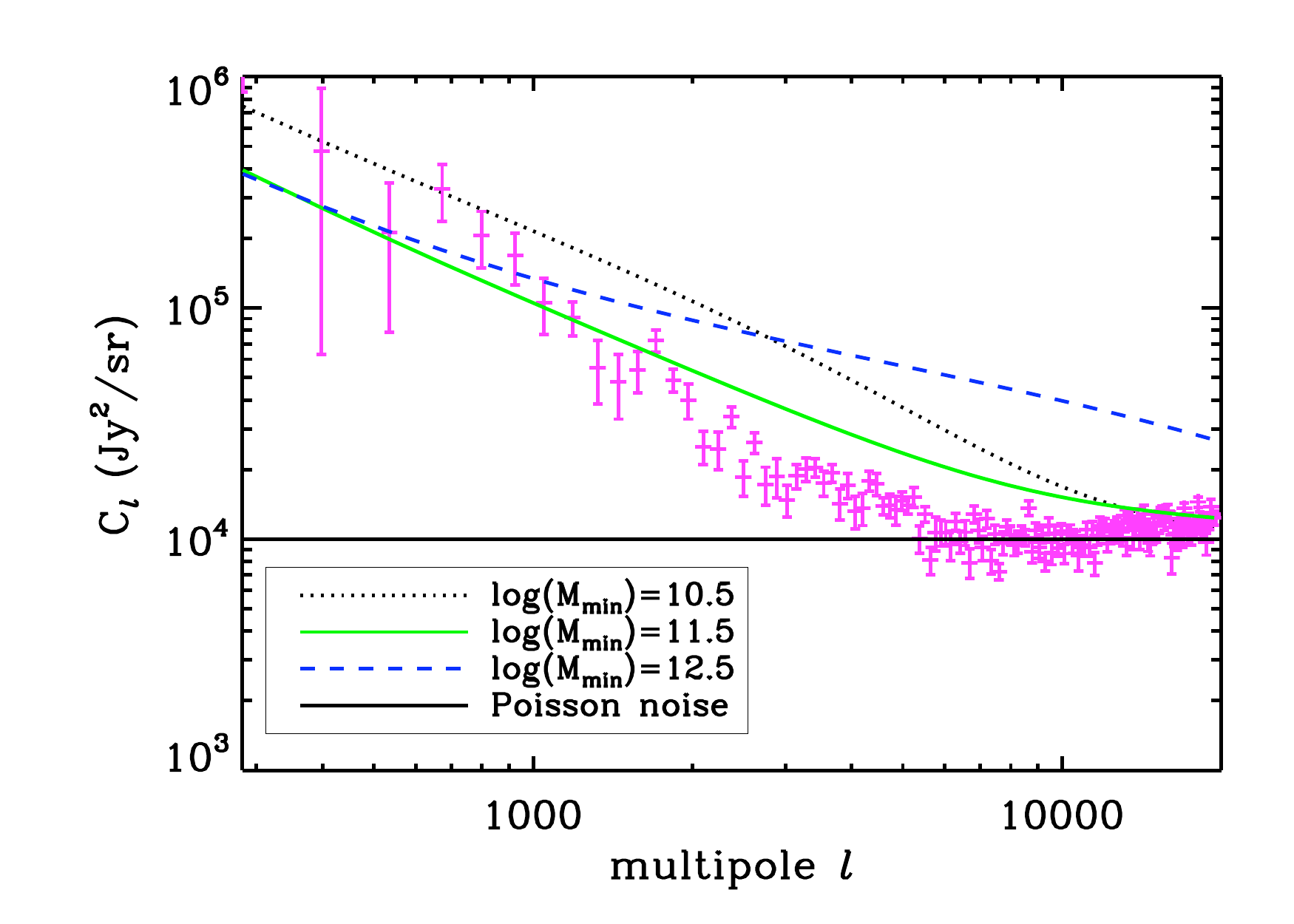}\includegraphics[scale=0.35]{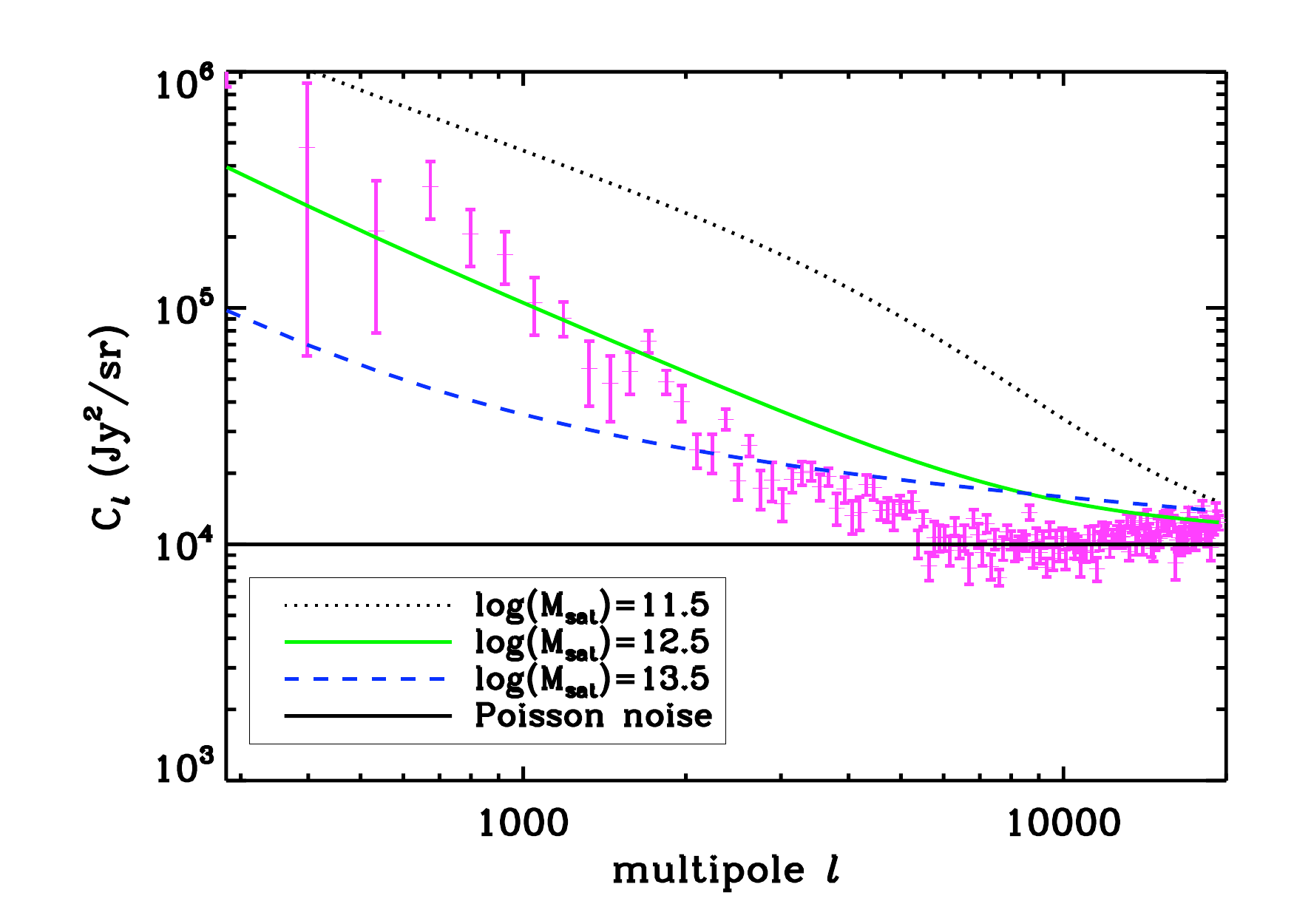}
\caption{CIBA power spectrum at 160~\um~obtained using several values of $\alpha_{sat}$,$M_{min}$ and $M_{sat}$. The level of the Poisson noise has also been added to the power spectra (black continous line). Pink dots are the data from \cite{2007ApJ...665L..89L} at 160~\um. When fixed, the parameters are those of the fiducial model, $\alpha_{sat} = 1.4$, $M_{min}=10^{11.5}M_{\Sun}$ and $M_{sat}=10^{12.5}M_{\Sun}$.
Left panel: The blue line is for $\alpha_{sat} = 1.8$, the green one for $\alpha_{sat} = 1.4$ and the black one for $\alpha_{sat} = 1$.
Middle panel: The blue line is the clustering power spectrum for $M_{min}=10^{12.5}M_{\Sun}$, the green one for $M_{min}=10^{11.5}M_{\Sun}$ and the black one for $M_{min}=10^{10.5}M_{\Sun}$.
Right panel: The blue line is for $M_{sat}=10^{13.5}M_{\Sun}$, the green one for $M_{sat}=10^{12.5}M_{\Sun}$ and the black one for $M_{sat}=10^{11.5}M_{\Sun}$. As expected, $C_\ell$s strongly depend on the halo parameters. We can expect strong degeneracies between those parameters.}
\label{fig:cl_var_hod}
\end{figure*}

\begin{figure}[!t]
  \includegraphics[width=\columnwidth]{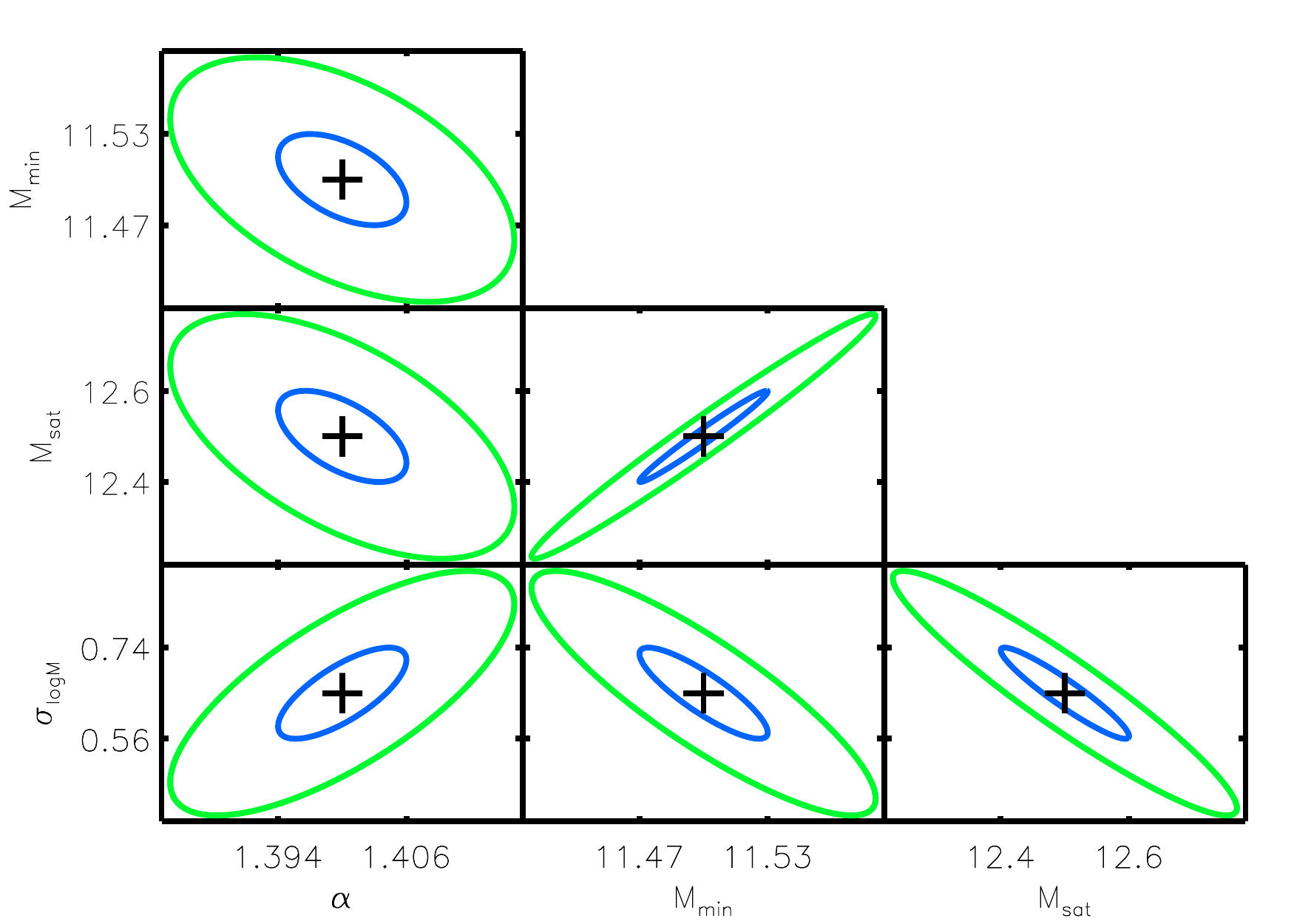}
  \caption{1$\sigma$ (blue) and 2$\sigma$ (green) likelihood contours of the halo model parameters computed with mock data from 100~$\mu$m to 1.3 mm.}
  \label{fig:ellipse_halo}
\end{figure}

\begin{figure*}[!t]
  \includegraphics[width=\textwidth]{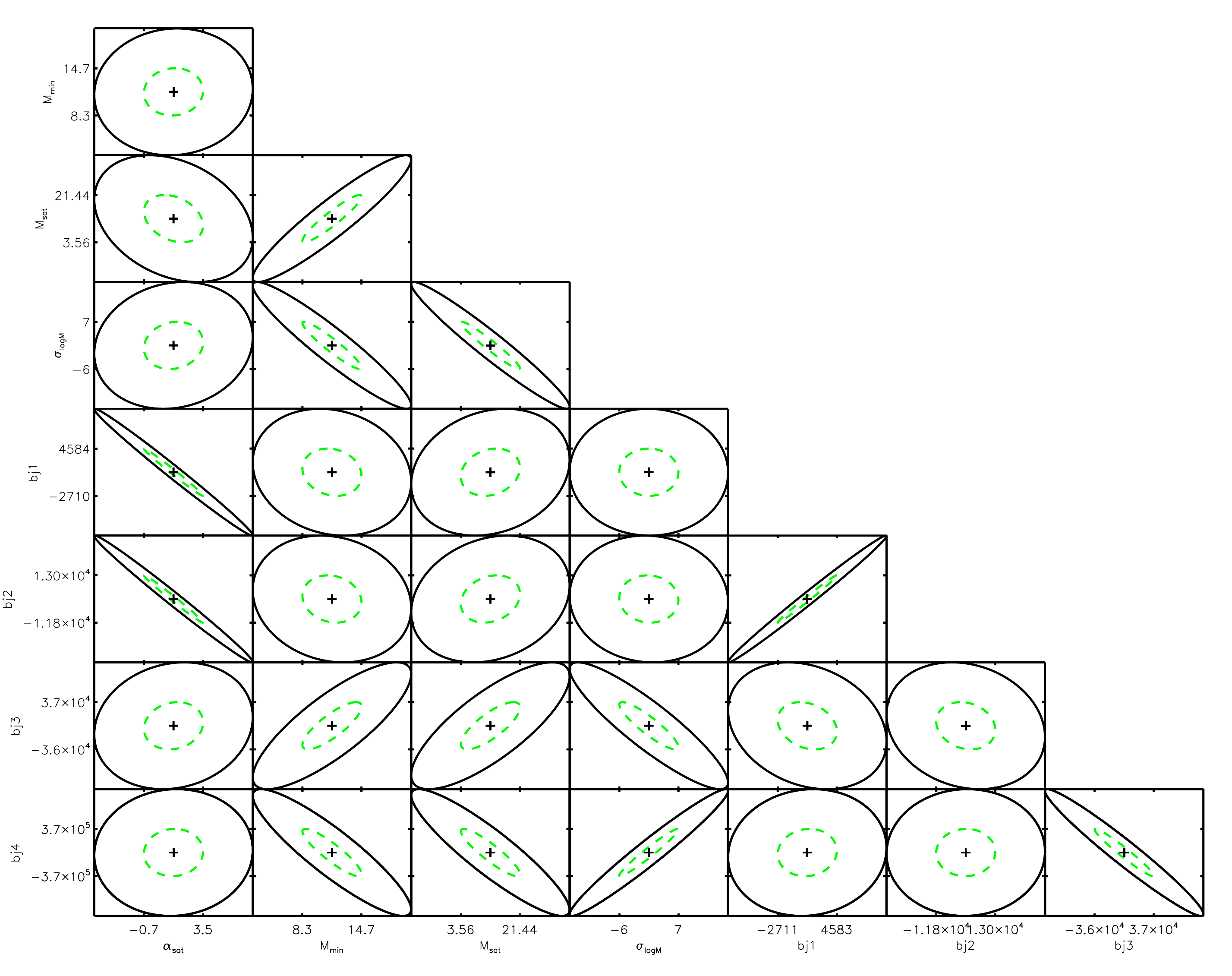}
  \caption{1$\sigma$ (green) and 2$\sigma$ (black) likelihood contours of the halo model parameters computed with mock data at 350~$\mu$m. Instead of using emissivities coming from the model of galaxies, we split \cl~in four redshift bins, on which we use the mean value of the emissivity on each bin $b_{j,i}$ that we consider as free parameters. The $b_{j,i}$ are hardly constrained with our data. $\alpha_{sat}$ is strongly degenerate with $b_{j,1}$ and $b_{j,2}$ which are the mean emissivities at low $z$. $M_{min}$ $M_{sat}$ and $\sigma_{logM}$ are degenerate with $b_{j,3}$ and $b_{j,4}$, the mean emissivities at high $z$.}
  \label{fig:ellipse_halo+bj}
\end{figure*}
The shape of the clustering power spectra strongly depends on the HOD parameters as shown on Fig. \ref{fig:cl_var_hod}. We vary $M_{min}$, $M_{sat}$ and $\alpha_{sat}$. Each panel shows the power spectrum at 160~$\mu$m measured by \citet{2007ApJ...665L..89L} in addition to the ones coming from the model. In each plot we vary only one HOD parameter and hold the others fixed to the values of the fiducial model, $\alpha_{sat} = 1.4$, $M_{min}=10^{11.5}M_{\Sun}$ and $M_{sat}=10^{12.5}M_{\Sun}$. Both the shape and amplitude strongly vary. The fact that some similar changes are observed using different parameters suggest strong degeneracies. We compute the Fisher matrix as in Sect.~\ref{par:gal_mod_param} and likelihood contours at 1- and 2-$\sigma$ are shown on Fig. \ref{fig:ellipse_halo}. The error bars on $\sigma_{logM}$ are very large, thus we fix its value to 0.65, following \citet{2010ApJ...719...88T} who studied the galaxy-clustering in optical surveys.  Using \cl~only we are not able to constrain its value. $M_{min}$ and $M_{sat}$ happen to be highly degenerated in the direction $M_{sat}=3.3M_{min}$. \\
In previous works using optically selected galaxies, $\alpha$ is usually set to 1 (\citet{2004MNRAS.355..819G}) and $M_{min}$ and $M_{sat}$ are the only parameters to be fitted to the data. Indeed, $\alpha$ is hardly constrained. For instance \citet{2010ApJ...719...88T} used the same halo distribution number on a sample of red and blue galaxies in the range $0.4<z<2$. They fitted well their correlation function fixing $\alpha=1$ and letting free $M_{min}$ and $M_{sat}$. But when they let $\alpha$ free in addition to the two others, they obtain unrealistic values for $\alpha$ \citep{2010ApJ...719...88T}. However, setting $\alpha=1$ might not be appropriate for CIBA. The halo model is commonly used in galaxies catalogs that are not deeper than $z\sim2$ and CIBA probe higher redshifts, especially at long wavelengths. Moreover optically-selected galaxies are not only star-forming galaxies, and there is no reason why optically-selected galaxies and star-forming galaxies should behave in the same way. \citet{2008MNRAS.383.1131M} used a similar form of the halo model to analyze the angular correlation function of 24~$\mu$m sources at $0.6<z<1.2$ and $z\geqslant1.6$. Using two halo density profiles (NFW and a steeper one $\rho\sim r^{-3}$), they derived $\alpha\sim0.7$ for the steeper profile and $\alpha\sim0.8$ for the NFW one. They get the same results for both sets of data. On the contrary, \citet{2010AA...518L..22C} computed the angular correlation function of sources detected at 250, 350 and 500~$\mu$m in Herschel/SPIRE data. They used the same halo distribution as ours and get $\alpha=1.3\pm0.4$, $\alpha<1.8$ and $\alpha<1.6$ at 250, 350 and 500~$\mu$m respectively. Finally, the \citet{2011arXiv1101.2028P} derived values of $\alpha$ compatible with 1. The discrepancy with \citet{2008MNRAS.383.1131M} may be due to the different properties of the bright galaxies selected at 24~$\mu$m and those that contribute to the CIBA at longer wavelengths. Here combining all CIBA measurements from 100 \um~to 1.3 mm and low to high multipoles, $\alpha$ is well constrained and it is not strongly degenerated with other parameters. Note that values of $\alpha>1$ implies that higher-mass halos contribute relatively more than smaller-mass ones compared to the halos in which lie optical galaxies such as those used by \citet{2010ApJ...719...88T}. \\
In our analysis of the degeneracies of the halo parameters we have only considered a set of parameters identical for all wavelengths which is not the case in reality. It could thus be that the degeneracies depend on wavelength. We therefore checked that the degeneracies were not significantly changing when we compute the Fisher matrix with various set of parameters corresponding to the wavelength best-fit models.\\
The halo parameters cannot be constrained by counts or LF as they only intervene in the clustering of galaxies in the equation of the \cl~(see Eq. \ref{eq:cl}). Therefore we cannot carry a joint analysis of the degeneracies of the halo parameters using counts/LF and \cl~data all together. Note that in principle, we could extrapolate the number count measurements to constrain the total number of galaxies, which also depends on the HOD parameters, but this would be a difficult measurement as it would be strongly dependent on the flux cut, for example.\\
Emissivities are given by the model of galaxy evolution but we want to investigate the degeneracies if they are binned in redshift and their values considered as free paramete,rs as in \citet{2011arXiv1101.1080A}. They carried this analysis at redshift between 0 and 4. In order to be coherent with what has been done previously and to take advantage of our redshift range, from 0 to 7, we split the whole redshift range in four bins, $0<z<0.9$, $0.9<z<2$, $2<z<3.5$, $3.5<z<7$. For each bin $i$ we take the mean value of the emissivity that we call $b_{j,i}$ with $i=\{1,2,3,4\}$ and we compute the \cl~at 350~\um~(we assume a combined Planck and Herschel power spectrum) and the associated Fisher matrix. Confidence levels are given on Fig.~\ref{fig:ellipse_halo+bj}. First we see that the halo occupation number degeneracies do not change much (see the previous paragraph) apart from the error bars which are much larger. $M_{min}$ is still strongly degenerated with $M_{sat}$ such as $M_{sat}= 2.7M_{min}$. The direction of the degeneracy is roughly the same as that derived using emissivities of the model as well as those of $\sigma_{logM}$ and $M_{sat}$ and $M_{min}$. Therefore, the degeneracy directions are all similar, using the emissivities or letting them free.\\
The degeneracies of the $b_j$ with the halo parameters depend on the redshift. Indeed, $\alpha_{sat}$ is highly degenerated with $b_{j1}$ ($0<z<0.9$) and $b_{j2}$ ($0.9<z<2$) and not at all with $b_{j3}$ ($2<z<3.5$) and $b_{j4}$ ($3.5<z<7$). Therefore $\alpha_{sat}$ is constrained by $z>2$ galaxies whereas the other halo parameters behave in the opposite way, they are not degenerate with $b_{j1}$ and $b_{j2}$ but with $b_{j3}$ and $b_{j4}$. This redshift dependency is emphasized by their own degeneracies. $b_{j1}$ and $b_{j2}$ are strongly correlated, the same is true for $b_{j3}$ and $b_{j4}$. The degeneracies using other wavelengths are only slightly different, we do not show them here. In general, the couples $(b_{j1},b_{j2})$ and $(b_{j3},b_{j4})$ are always strongly degenerate and the $b_{j,i}$ are degenerate with the halo parameters as shown on Fig.~\ref{fig:ellipse_halo+bj}. Such a degeneracy dependence with the redshift has to be still understood. \\
In order to compare our results with those of \citet{2011arXiv1101.1080A}, we carry a similar analysis using their redshift bins, that is to say, $0<z<1$, $1<z<2$, $2<z<3$ and $3<z<4$. We observe the same behavior as described previously but different from their results. They used Monte Carlo Markov Chains to compute the degeneracies and usually the two dimensional probability distributions have two peaks (see their fig. S 13). Their $S_i$ are equivalent to our $b_{j,i}$. $S_1$ is degenerate with the three others $S_i$, whereas there is no degeneracy between $S_3$ and $S_4$. We do not discuss the degeneracy of the halo parameters with $S_i$ as their parameterization of the halo occupation number slightly differ from ours.\\
Using $b_{j,i}$ and fitting them on the data avoids us to rely on a model of evolution of galaxies. However they are poorly constrained with the present data. Moreover the degeneracies between the $b_{j,i}$ and the halo parameters strongly depend on the halo parameterization used. 
\section{Interpreting measurements}

Now that we have determined a model and the associated parameters degeneracies,  we discuss their physical interpretation.

\subsection{Redshift and halo-masses contribution to the power spectrum}

\begin{figure}[!ht]
  \includegraphics[height=0.9\textheight]{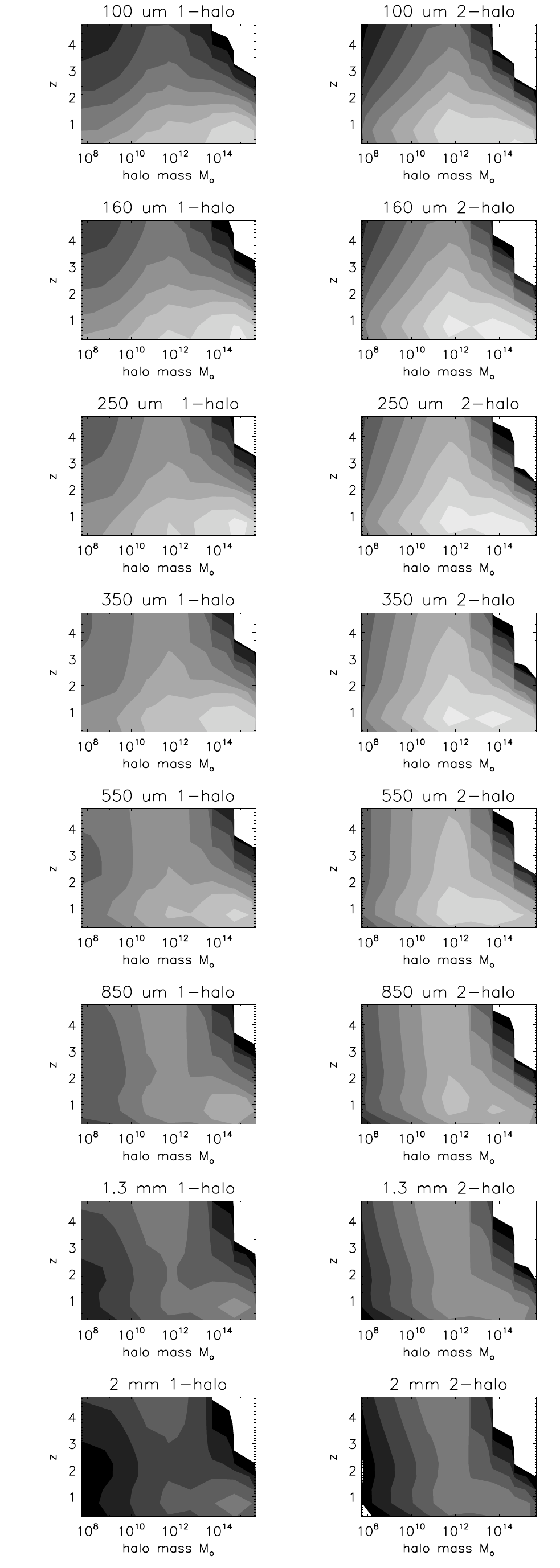}
  \caption{Contribution of halo masses and redshift to the \cl~from 100~$\mu$m to 2 mm. The first column shows the redshift and mass contribution to the 1-halo term ($\ell=2002$) and the second column represents the same contributions to the 2-halo term ($\ell=100$). The light grey corresponds to the highest contribution to the \cl s. The step of the color range is logarithmic and the scale is the same for both columns. The 1-halo term is dominated by high masses at low $z$ at all wavelength. At all wavelength, the 2-halo term is dominated by a large range of masses at low $z$ and by intermediate masses at higher $z$. The 1- and 2-halo term are sensitive to different mass regimes. The former to high mass halos and the latter to intermediate mass halos.}
  \label{fig:cl_contrib_zm}
\end{figure}

\begin{figure}[!ht]
  \includegraphics[width=\columnwidth]{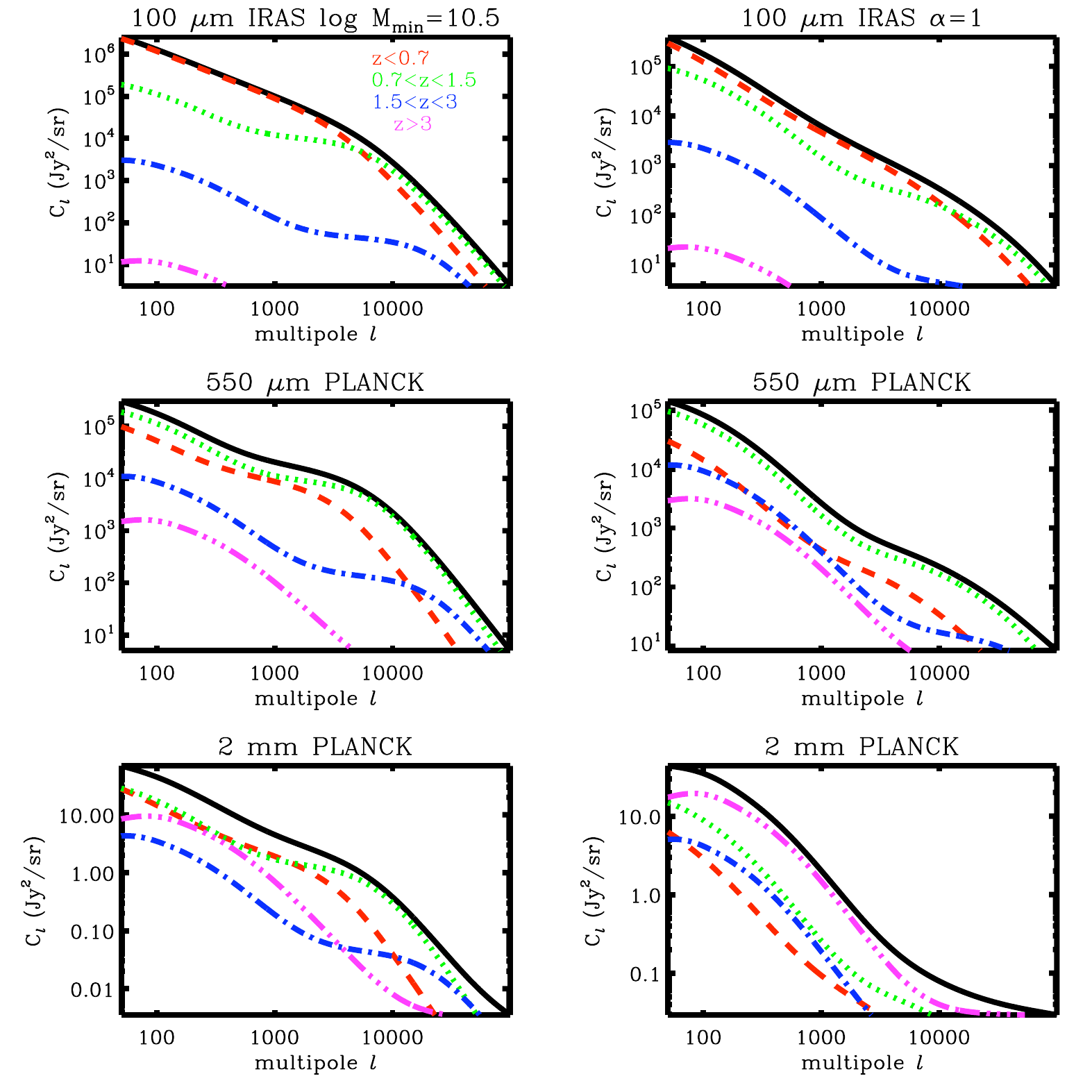}
  \caption{Redshift distribution for several wavelengths for halo parameters different from the values of the fiducial model. The left panels are for $\log M_{min}=10.5$ instead of $\log M_{min}=11.5$ and the right panels are for $\alpha= 1$ instead of $\alpha_{sat} = 1.4$. The redshift distribution depends strongly on the chosen halo parameters although the same trend is observed~: the low redshift dominate at short wavelengths, this contribution decreases with the wavelength and as the wavelength gets longer the contribution of high redshifts increases.}
  \label{fig:compare_cl_param}
\end{figure}

The left panel of Fig. \ref{fig:compare_cl_distrib_z} shows the contribution to the \cl~by several redshift bins. As stated previously, the shorter the wavelength, the more important is the relative contribution from the low redshift. For example, while $z<0.7$ contributes significantly at 100 and 160~$\mu$m, it becomes much less important in the millimeter range. Reversely, the high redshift bin ($z>3$) is negligible at short wavelength but has an increasing contribution when the wavelength increases. The redshift distribution can change with the choice of the halo parameters as shown on Fig. \ref{fig:compare_cl_param}. We have changed the values of $\alpha_{sat}$ on the left panels and that of $M_{min}$ on the right panels (the change is in the 1$\sigma$ error bars of the best fit found by \citet{2011arXiv1101.2028P}), the other parameters are those of the fiducial model. We recover the trend noticed above more or less emphasized. \\
Further, not only the redshift of the galaxies probed depends on the wavelength, but so does the mass of the halos in which they are embedded. Fig.~\ref{fig:cl_contrib_zm} shows the contribution of mass and redshift to the 1- and 2-halo terms (at $\ell=2002$ and $\ell=100$ respectively) from 100~$\mu$m to 2 mm. High-mass halos ($M>10^{13}M_{\odot}$) contribute the more to the 1-halo term from 100~$\mu$m to 2 mm at low redshift, and this dominant mass range stays constant with wavelength. More massive halos contain more galaxies than smaller ones therefore the galaxies contained in those halos contribute more to the angular power spectrum. This can be explained qualitatively in the following way. According to the mass function, at say $z=0.5$, there are one thousand times more halos of $M=10^{11}M_\odot/h$ than halos of mass $M=10^{14}M_\odot/h$. According to the HOD, one out of one hundred $10^{11}M_\odot/h$ halo hosts a galaxy, whereas $M=10^{14}M_\odot/h$ mass halos hosts on average 10 galaxies. Since the contribution to the 1-halo terms goes like $N_{gal}^2$, lore massive halos contribute relatively more to the 1-halo term. Note that this trend is less prononced for the 2-halo term since it goes like $N_{gal}$. At all wavelengths, as the redshift increases, the dominant mass range decreases to $M\sim10^{11-13}M_{\odot}$ as halos at higher redshifts are smaller than those at $z=0$. \\
 The 2-halo term does not exhibit the same behavior~: at short wavelengths, halos in a large range of mass $10^{11}-10^{15}M_{\odot}$ at low $z$ contribute to the power spectrum. In parallel, intermediate masses contribute at higher $z$. As the wavelength increases, the relative contribution between high mass at low $z$ and intermediate mass at high $z$ becomes close to unity. It ends up in an equal contribution from high mass at low $z$ and from intermediate mass at high $z$ at 2 mm.  Intermediate mass halos are more abundant which explains their high contribution. Both the 1 and 2-halo terms are sensitive to different mass regimes which evolve with the wavelength and thus with the redshift.\\
\cite{2008MNRAS.383.1131M} selected 24~$\mu$m sources at $1.5<z<3$ and deduced from their correlation function that they lie in $10^{13}M_{\odot}$ halos. \cite{2009ApJ...707.1766V} found an $M_{eff}\sim10^{13.2}M_{\odot}$ for unresolved galaxies at $z>1$ at 250, 350 and 500~$\mu$m. \cite{2008ApJ...687L..65B} derived the angular autocorrelation function of dust-obscured galaxies selected with a color criterion. They determined that they are in halos with an average mass of $10^{12.2}M_{\odot}$. \cite{2007AA...475...83G} selected star forming galaxies at 24~$\mu$m, derived the projected correlation function and found that LIRGs lie in halos $M>3\times10^{13}M_{\odot}$. All these results are in agreement with ours. Overall, halos with masses such as $10^{12-13}M_{\odot}$ contribute the most to power spectra at all redshifts. However, we want to reemphasize here that these conclusions are model-dependent and depend on the particular emissivity model used, as discussed before. This is particularly true for the higher-$z$ contribution, say $z>2$.\\

\subsection{Linear bias}
\begin{figure}[!t]
  \includegraphics[width=\columnwidth]{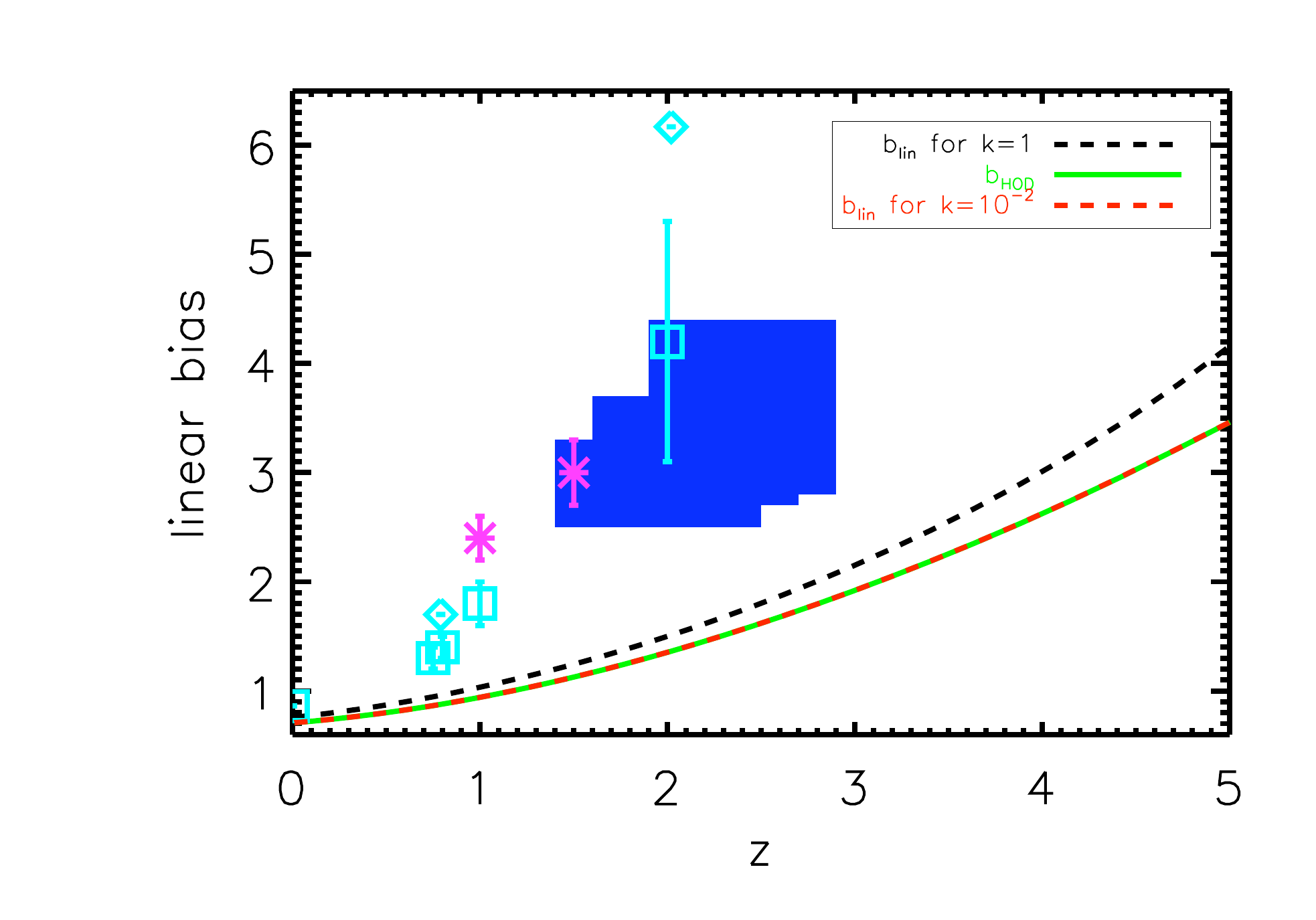}
  \caption{Linear bias for several values of $k$ (in (Mpc/$h$)$^{-1}$) and HOD bias. Light blue squares and diamonds bias values from resolved galaxies and pink crosses represent biases from unresolved galaxies measurements (see Table \ref{tab:bias_list}). The big blue squares represent the HOD biases coming from resolved galaxies from \citet{2010AA...518L..22C}. Light blue diamonds and squares show HOD and linear biases, respectively.}
  \label{fig:plot_bias}
\end{figure}
Within our halo model we derive the linear bias as a function of the redshift following:
\be 
b_{lin}(k,z)=\sqrt{\frac{P_{gg}(k,z)}{P_{lin}(k,z)}}
\ee
where $P_{gg}(k,z)$ is the galaxy-galaxy power spectrum coming from our model, $P_{lin}(k,z)$ is the linear DM power spectrum and $b_{lin}(k,z)$ the linear bias.\\
We will call the effective bias coming from the HOD model, $b_{HOD}$. On large scale $u(k\rightarrow 0,M)\sim1$, so the HOD bias from Eq. \ref{eq:p12h} is~:
\be
b_{HOD}(z) = \int{dM\frac{dN}{dM}b(M)\frac{<N_{gal}>}{\bar{n}_{gal}}}
\ee
 We plot in Fig. \ref{fig:plot_bias} the linear biases and the HOD bias for our fiducial model as a function of redshift for $k=1$ (Mpc/$h$)$^{-1}$ and $k=10^{-2}$ (Mpc/$h$)$^{-1}$ where we also add current measurements detailed in Tab. \ref{tab:bias_list}. Note that in the linear regime, the HOD and linear biases are identical which is the case at $k=10^{-2}$ (Mpc/$h$)$^{-1}$ but not at $k=1$ (Mpc/$h$)$^{-1}$. On small spatial scales, the $u(k\rightarrow 0,M)\sim1$ is not true therefore it is strongly different from the linear bias for $k=1$ as shown on Fig. \ref{fig:plot_bias}.\\
For both scales, our linear biases as well as the HOD bias is not in agreement with measurements. Neither of them show the same trend as the data points. The measured linear biases as well as the HOD biases grow quicker towards higher values than the biases extracted from our model.\\
HOD bias measurements are from different HOD. \cite{2010AA...518L..22C} used the same halo occupation number to fit the correlation function and they found different parameters than ours. When using their parameters set, we do recover their results. The discrepancy could be explained by the fact that these measurements result from correlation function analysis, thus from resolved sources which is not the population we are studying here.\\
Concerning the linear bias determined with unresolved galaxies \citep{2007ApJ...665L..89L,2009ApJ...707.1766V}, our linear bias is not in agreement with the measurements either. Indeed when using unresolved sources, the determination of the bias requires the use of emissivities, which are strongly model dependent (see Sect. \ref{par:emissivities}) and it can affect the bias. 

\begin{table*}\centering
\begin{tabular}{*{6}{c}}
\hline\hline
wavelength     &kind of galaxies   &reference                     &$<z>$               &$b_{HOD}$                   &$b_{lin}$ \\
\hline 
24             &resolved           &\citet{2008MNRAS.383.1131M}    & 0.79               & 1.70                       &\\
24             &resolved           &\citet{2008MNRAS.383.1131M}    & 2.02               & 6.17                       &\\
24             &resolved           &\citet{2008ApJ...687L..65B}    & 2                  &                            & 3.1-5.3\\
24             &resolved           &\citet{2007AA...475...83G}     & 0.75               &                            & 1.3$\pm$0.1\\
24             &resolved           &\citet{2007AA...475...83G}     & 0.8                &                            & 1.4$\pm$0.1\\
24             &resolved           &\citet{2007AA...475...83G}     & 1                  &                            & 1.8$\pm$0.2\\
100            &resolved           &\citet{1992MNRAS.258..134S}    & 0                  &                            & 0.86\\
160            &background         &\citet{2007ApJ...665L..89L}    & 1                  &                            & 2.4$\pm$0.2\\
250-350-500    &background         &\citet{2009ApJ...707.1766V}    & 1.5                & 2.2$\pm$0.2                & 3$\pm$0.2\\
250            &resolved           &\citet{2010AA...518L..22C}     &$ 2.1^{+0.4}_{-0.7}$  & 2.9$\pm$0.4                &\\
350            &resolved           &\citet{2010AA...518L..22C}     &$ 2.3^{+0.4}_{-0.7}$  & 3.2$\pm$0.5                &\\
250            &resolved           &\citet{2010AA...518L..22C}     &$ 2.6^{+0.3}_{-0.7}$  & 3.6$\pm$0.8                &\\
\hline
\end{tabular}
\caption{Linear and effective bias measurements. The third column gives the mean redshift of the galaxies probed and the last but one lists the HOD/effective bias values and the last one gives the linear bias.}
\label{tab:bias_list}
\end{table*}

\subsection{Influence of the mean emissivities}\label{par:emissivities}
\begin{figure}[!t]
\includegraphics[width=\columnwidth]{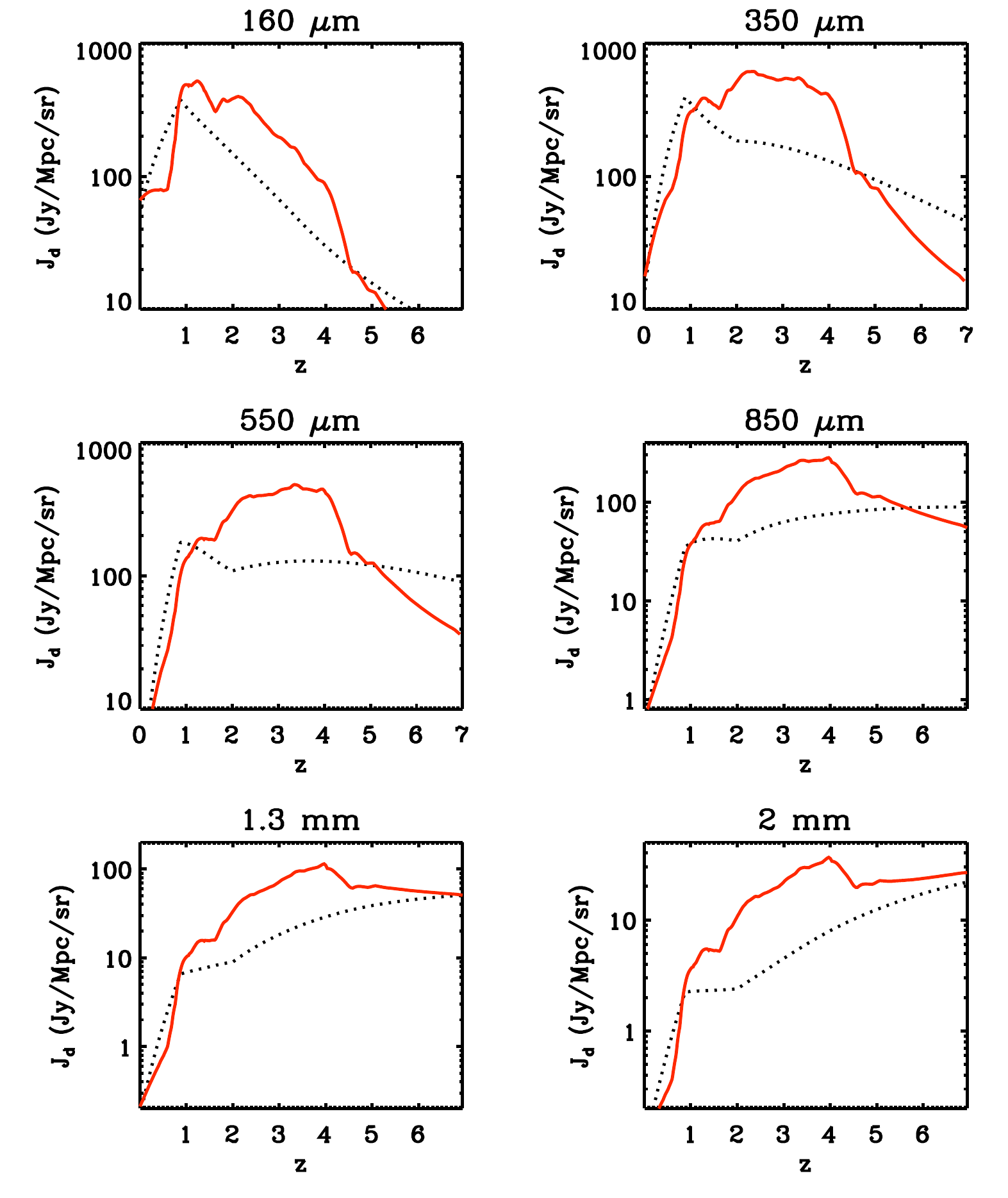}
\caption{Emissivities used by \cite{2009ApJ...707.1766V} coming from \cite{2004ApJS..154..112L} (red continous line) and ours (black dotted line) coming from \citet{2011A&A...529A...4B} at several wavelengths.}
\label{fig:plot_compare_jd}
\end{figure}

\begin{figure}[!h]
\includegraphics[width=\columnwidth]{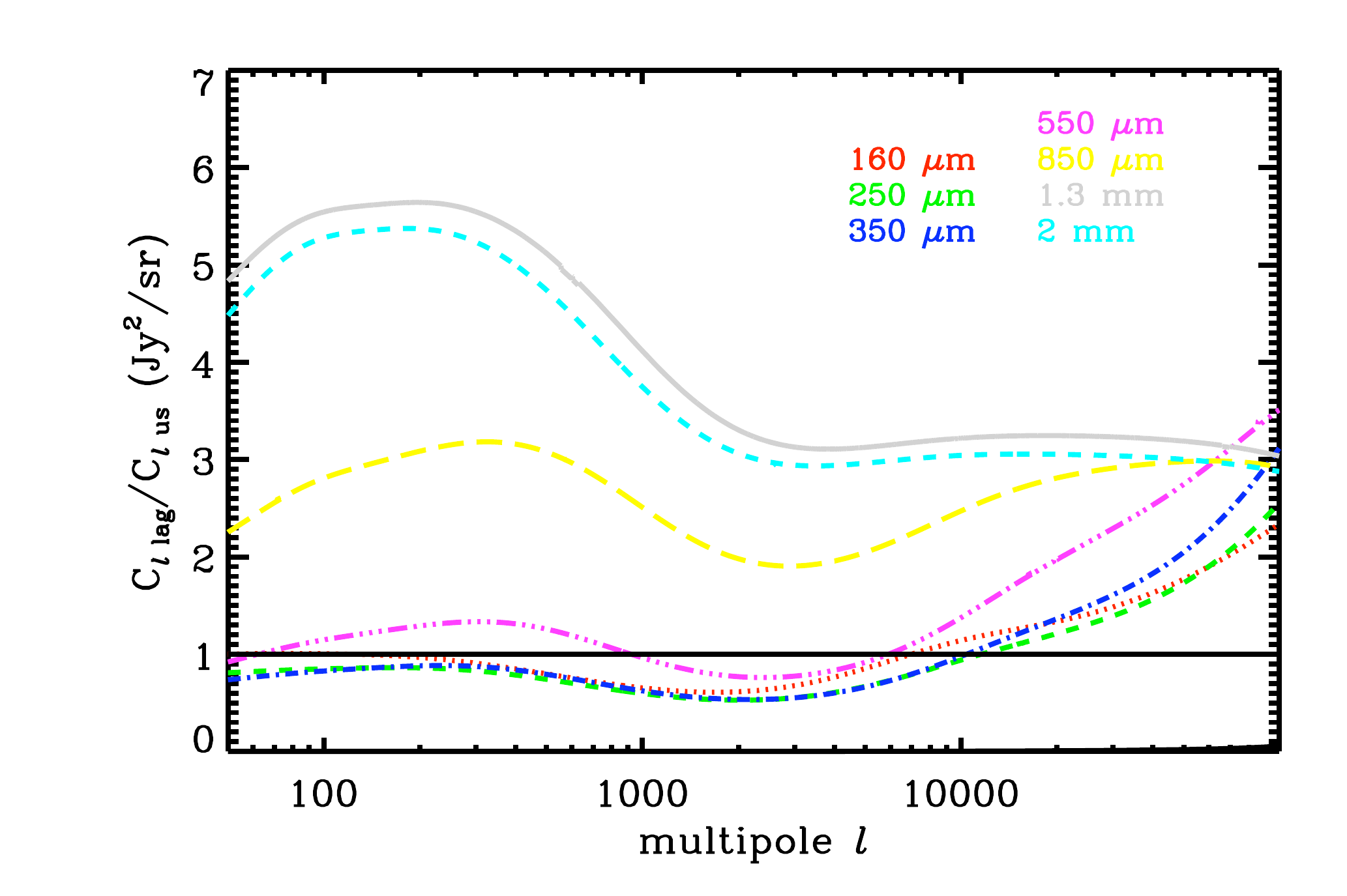}
\caption{Power spectra ratios computed using the emissivities of \cite{2004ApJS..154..112L} ($C_{\ell,lag}$)  and \citet{2011A&A...529A...4B} ($C_{\ell,us}$), for several wavelengths.}
\label{fig:compare_cl}
\end{figure}

\begin{figure*}[!t]
\includegraphics[width=1.5\columnwidth]{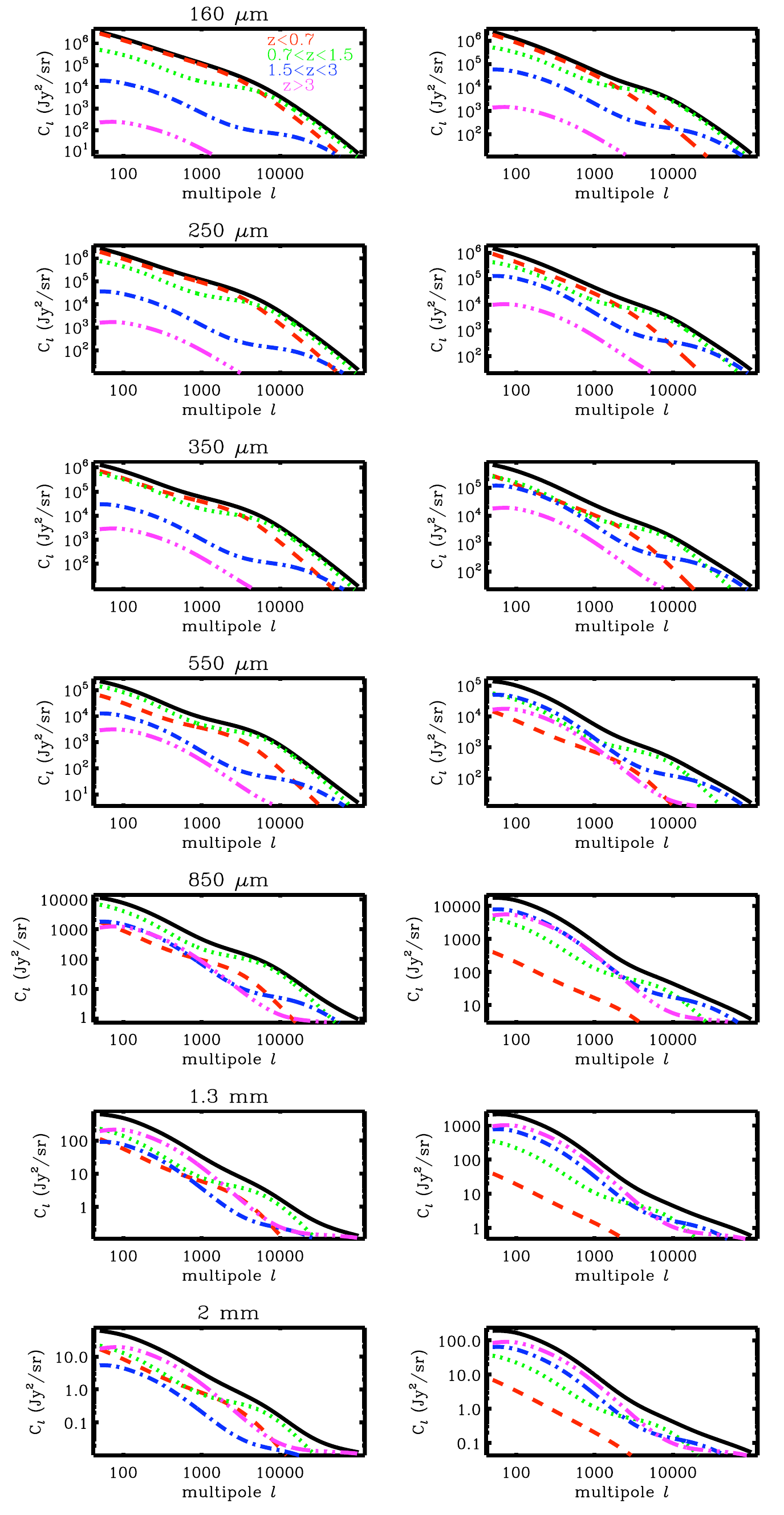}\centering
\caption{Redshift contribution to the \cl~at several wavelengths using our emissivities on the left and using \cite{2004ApJS..154..112L} emissivity on the right. In both cases, the contribution of high redshift increases with the wavelength but the evolution goes faster with \cite{2004ApJS..154..112L} emissivities. These redshift distributions are strongly varying with the  halo parameters (see Fig. 10)}
\label{fig:compare_cl_distrib_z}
\end{figure*}

Previous models such as those of \cite{2007ApJ...665L..89L} and \cite{2009ApJ...707.1766V} have used emissivities coming from \cite{2004ApJS..154..112L}. In Fig.~\ref{fig:plot_compare_jd}, we plot the emissivity used in this paper as well as the \citet{2004ApJS..154..112L} ones for reference. The peak at $z\sim1$ in our emissivities is due to the parameterization of the LF. Despite the shapes of the emissivities of \cite{2004ApJS..154..112L} and ours are different, they display similar trends. The relative contributions of high redshifts increases with wavelength while the contribution of low redshifts decreases. According to \cite{2011A&A...525A..52J}, the model of \cite{2004ApJS..154..112L} predicts too much power at high $z$. As the latter is forced to reproduce levels of the CIB and number counts, it does not predict enough power at low $z$. Therefore it predicts more galaxies at high $z$ and less at low $z$. To illustrate how it influences our results, we show on Fig. \ref{fig:compare_cl} the ratios of the power spectra computed with our emissivities and those from \cite{2004ApJS..154..112L}. Up to 550~\um~the ratio is around 1 up to $\ell\sim10000$ and it increases strongly at higher $\ell$. At longer wavelength, the difference is much larger. It is in line with the over-prediction of power at high redshift of the model of \cite{2004ApJS..154..112L}. The same halo parameters have been used for this plot, however it is clear that when fitting the model to the data with both emissivities we will not find the same halo parameters. \\
As said previously the \cite{2004ApJS..154..112L} model predicts too much power at high $z$, thus we need to compare the contribution in redshift to the \cl. They are given on Fig. \ref{fig:compare_cl_distrib_z}. In both cases we observe the same trend~: high redshift contribute more and more as the wavelength increases. However, with \cite{2004ApJS..154..112L} emissivities this evolution goes faster. For instance, at 250~\um, the contribution of $1.5<z<3$ galaxies is of the same order of magnitude that those of $0.7<z<1.5$ whereas in our case the former is more than one order of magnitude inferior to the latter. As we go towards longer wavelengths, the two highest redshift bins have an increasing contribution, and both dominate the power spectrum at 850~$\mu$m using \cite{2004ApJS..154..112L} emissivities whereas using our emissivities $0.7<z<1.5$ galaxies contribute also strongly. At 1.3 and 2 mm both the highest redshift bins contribute the most but in our case, only the highest redshift bin dominate and the $1.5<z<3$ bin has a smaller contribution. Therefore the shape of the emissivities strongly influences our results, parameters determination and redshift distribution. The interpretation of clustering measurements is thus based on the use of a reliable model of evolution of galaxies.

\subsection{Contribution of LIRGs and ULIRGs}
\begin{figure*}
\includegraphics[width=\textwidth]{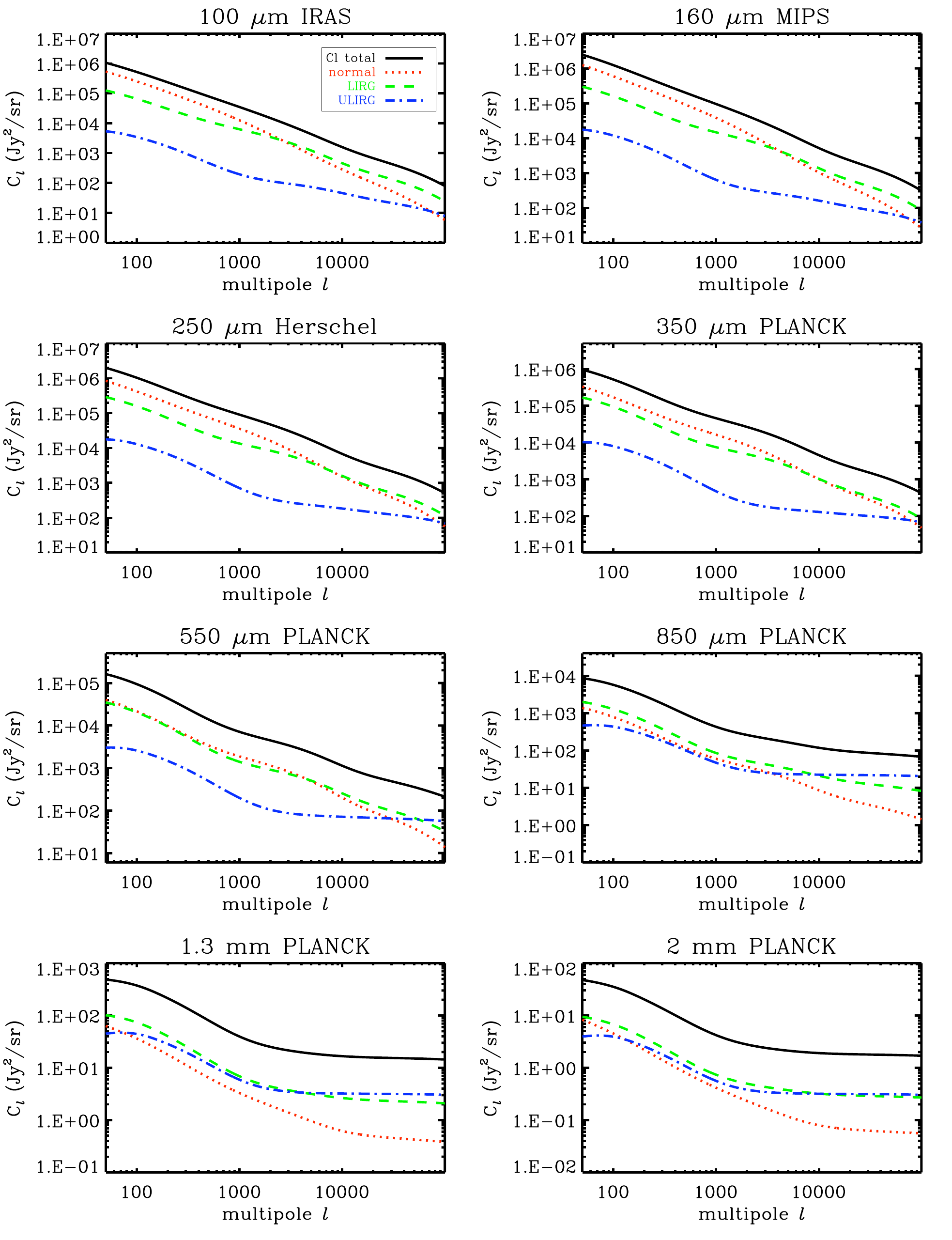}
\caption{Contribution to the \cl~of normal galaxies, LIRGs and ULIRGs at several wavelengths. When computing power spectra by splitting the contributions of normal galaxies, LIRGs and ULIRGs, cross terms appear. We do not not show them on that plot, therefore the sum of the three power spectra do not make the total. We recover that normal and LIRGs dominate at short wavelengths, thus at small redshifts. As the wavelength increases, and at the mean time the high-redshift contribution, the relative contribution of ULIRGs increases.}
\label{fig:contrib_lirg}
\end{figure*}

%---------------------------------------------
Star-forming galaxies are split in several categories according to their luminosities. Normal, luminous infrared, and ultra luminous infrared galaxies have luminosities $L_{IR}<10^{11}M_{\odot}$, $10^{11}M_{\odot}<L_{IR}<10^{12}M_{\odot}$, $L_{IR}>10^{12}M_{\odot}$, respectively. LIRGs dominate the infrared energy output at $z\sim1$ and ULIRGs at $z\sim2$ \citep{2005ApJ...630...82P,2005ApJ...632..169L,2007ApJ...660...97C} therefore we look at their contribution to the \cl~and to their evolution with the wavelength. To do so we split the emissivities in the following way~:
\be 
\bar{j}_{\lambda} = \bar{j}_{\lambda}^{normal}+\bar{j}_{\lambda}^{LIRG}+\bar{j}_{\lambda}^{ULIRG}
\ee
and this contribution is squared in the \cl. Therefore cross terms appear~: 
\bea 
C_{\ell,total} &=& C_{\ell,Normal}+C_{\ell,LIRG}+C_{\ell,ULIRG}+2(\times C_{\ell,Normal/LIRG}\nonumber\\
&+&2 C_{\ell,Normal/ULIRG}+ C_{\ell,LIRG/ULIRG})
\eea
We plot in Fig. \ref{fig:contrib_lirg} the contributions of normal, LIRGs and ULIRGs. Note that the sum of the three contributions does not make the total power spectrum because the cross terms are not shown. \\
Normal galaxies and LIRGs both dominate the power spectrum up to 550~\um. The contribution of LIRGs increases slightly and finally dominates from 850~\um~to 2 mm. ULIRGs never clearly dominate the power spectrum at long wavelength, however their relative contribution increases at long wavelengths, from 850~\um~to 2 mm.\\
Therefore, we do recover what is expected from previous works. Normal galaxies dominate at low redshift, LIRGs at $z\sim1$ and ULIRGs contribute in the same way as the others at high redshift and thus at long wavelengths.

\section{Conclusion}
%---------------------------------------------

We presented a new model of the clustering of star-forming galaxies in the Cosmic Infrared Background anisotropies. We interfaced a parametric model of star-forming galaxies evolution with a halo distribution approach. The model is fully parametric. Fixing the cosmology, it depends on the parameters of the model of galaxies and the HOD. We computed power spectra from 100~$\mu$m to 2 mm for IRAS, Spitzer/MIPS, Herschel/SPIRE and Planck/HFI spectral bands. We showed how power spectra can depend on the parameters and we concluded that the parameters of the model of galaxies can hardly be constrained using \cl~only. Number counts and luminosity functions data are required. Fixing them at the mean values found by \cite{2011A&A...529A...4B}, we explored the HOD parameters constraints and degeneracies. The combination of \cl~and counts/LF data do not break the degeneracies but constraints are slightly improved. Some of the parameters are strongly degenerate, especially $M_{min}$ and $M_{sat}$ with $M_{sat}=xM_{min}$ with $x\sim3$ where $x$ is the direction of the degeneracy. \\
 We have shown that the 1-halo term can be detected at all wavelengths and that galaxies at high redshift lie in smaller halos than those at lower redshift. The level of the shot noise might not be reached with certain instruments such as Planck because of their angular resolution. However, this does not apply to the South Pole Telescope and to Herschel as they have a better angular resolution. \\
Using our fiducial model, we computed the halo mass and redshift contribution to the power spectra. Higher redshift galaxies contribute more at long wavelengths. Not surprisingly, the 1- and 2- halo terms do not have the same mass dependence. We found that high mass halos contribute the most to the 1 halo term whereas the 2-halo term is dominated by intermediate mass halos which are most numerous. Our model strongly depends on the emissivity given by the evolution model of galaxies, and we compare the resulting \cl s with those obtained using the emissivities coming from \cite{2004ApJS..154..112L}. We have shown that the halo parameters strongly depend on the emissivities when data are fitted.\\ 
In order to avoid the use of a model of evolution of galaxies, we have split the redshift range in four bins and computed the \cl~using the mean emissivity on these four redshift bins as in \citet{2011arXiv1101.1080A}. We considered these four parameters as free. They are not very well constrained, they cannot give any constraints on models of galaxies.\\
We investigate the contribution of LIRGs and ULIRGs to the power spectra and its evolution with the wavelength. Our results are in agreement with previous studies of normal galaxies, LIRGs and ULIRGs contribution to the CIB and to the luminosity functions. Normal galaxies dominate the power spectrum at low redshift. As the redshift increases, LIRGs dominate \cl. Meanwhile, the contribution of ULIRGs keeps increasing up to 2 mm. \\
The main unknown in CIB anisotropies power spectrum measurements are the redshift distributions of CIB galaxies. The coming results from Planck and Herschel will enable a great leap in the understanding of the clustering of star-forming galaxies and its redshift evolution, measuring the cross power spectra between wavelengths. These new measurements will help to break some degeneracies and will allow to make more precise measurements of the star formation density at high redshift, and the characteristic mass of the dark matter halo at which the efficiency of the star formation is maximal.

\begin{acknowledgements}
Part of the research described in this paper was carried out at the Jet Propulsion Laboratory, California Institute of Technology, under a contract with the National Aeronautics and Space Administration. The authors would like to thank Mathieu Langer for very useful comments that improved this manuscript.
\end{acknowledgements}

\bibliographystyle{aa}

\bibliography{Biblio}

\end{document}